\documentclass[fleqn,usenatbib]{mnras}

\usepackage{newtxtext}
\usepackage{amsmath}
\usepackage{MnSymbol}
\usepackage{enumitem}
\usepackage{cancel}
\usepackage[T1]{fontenc}

\newcommand{\hp}{H^{+}}
\newcommand{\hep}{He^{+}}
\newcommand{\hepp}{He^{+2}}
\newcommand{\opp}{O^{+2}}
\newcommand{\oppp}{O^{+3}}
\newcommand{\cpp}{C^{+2}}
\newcommand{\cppp}{C^{+3}}
\newcommand{\npp}{N^{+2}}
\newcommand{\sippp}{Si^{+3}}
\newcommand{\fep}{Fe^{+}}
\newcommand{\hei}{He~{\sc i}}
\newcommand{\heii}{He~{\sc ii}}
\newcommand{\ciii}{C~{\sc iii]}}
\newcommand{\civ}{C~{\sc iv}}
\newcommand{\nii}{[N~{\sc ii]}}
\newcommand{\niii}{N~{\sc iii]}}
\newcommand{\niv}{N~{\sc iv]}}
\newcommand{\nv}{N~{\sc v}}
\newcommand{\oiiis}{O~{\sc iii]}}
\newcommand{\oiii}{[O~{\sc iii]}}
\newcommand{\oiv}{O~{\sc iv]}}
\newcommand{\ovi}{O~{\sc vi}}
\newcommand{\siiv}{Si~{\sc iv}}
\newcommand{\feii}{Fe~{\sc ii}}
\newcommand{\mgii}{Mg~{\sc ii}}
\newcommand{\femg}{Fe~{\sc ii}/Mg~{\sc ii}}

\newcommand{\mghb}{Mg~{\sc ii}/H$\beta$}
\newcommand{\nagao}{(Si~{\sc iv}+O~{\sc iv]})/C~{\sc iv}}
\newcommand{\nvciv}{N~{\sc v}/C~{\sc iv}}
\newcommand{\nivciv}{N~{\sc iv]}/C~{\sc iv}}
\newcommand{\ciiiciv}{C~{\sc iii]}/C~{\sc iv}}
\newcommand{\oiiiciii}{O~{\sc iii]}/C~{\sc iii]}}
\newcommand{\oiiiciv}{O~{\sc iii]}/C~{\sc iv}}
\newcommand{\niiioiii}{N~{\sc iii]}/O~{\sc iii]}}
\newcommand{\feviinev}{[Fe~{\sc vii]}/[Ne~{\sc v]}}

\usepackage{CJKutf8}
\usepackage{amsmath,graphicx,longtable,hyperref}
\usepackage{natbib,threeparttable,rotating,placeins}
\usepackage{ulem,changepage}
\usepackage{xcolor}
\hypersetup{
    colorlinks,
    linkcolor={red!50!black},
    citecolor={blue!50!black},
    urlcolor={blue!80!black}
}

\title[Metal Enrichment in AGN Disks]{Metal Enrichment due to Embedded Stars in AGN Discs}

\author[J. Huang et al.]{Jiamu Huang (黄嘉沐)$^{1,2}$ \thanks{E-mail: jiamu$\_$huang@ucsb.edu (JH); lin@ucolick.org (DL); shields@lfastro.org (GS)}, 
Douglas N. C. Lin (林潮)$^{1, 3}$,
Gregory Shields$^{4}$
\\
$^{1}$Department of Astronomy and Astrophysics, University of California, Santa Cruz, CA 95064, USA\\
$^{2}$Department of Physics, University of California, Santa Barbara, CA 93106, USA\\
$^{3}$Institute for Advanced Studies, Tsinghua University, Beijing, 100086, China\\
$^{4}$Laguna Falls Institute for Astrophysics, 2001 Big Canyon Drive, Austin, TX 78746
}

\date{Accepted 2023 August 26. Received 2023 August 25; in original form 2023 April 2}
\pubyear{2022}
\begin{document}
\label{firstpage}

\begin{CJK*}{UTF8}{gbsn}

\pagerange{\pageref{firstpage}--\pageref{lastpage}} 

\maketitle
\end{CJK*}

\begin{abstract}
We separately assess elemental abundances in AGNs' broad and narrow emission 
line regions (BLR and NLR), based on a critical assessment of published results together 
with new photoionization models.  We find 1) He/H enhancements in some AGN, exceeding 
what can be explained by normal chemical evolution and confirm 2) super-solar $\alpha$ abundance, 
though to a lesser degree than previously reported.  We also reaffirm 3) a N/O ratio consistent 
with secondary production; 4) solar or slightly sub-solar Fe abundance;  and
5) red-shift independent metallicity, in contrast with galactic chemical evolution.
We interpret 6) the larger metallicity in the BLR than NRL in terms of an {\it in situ} 
stellar evolution and pollution in AGN discs (SEPAD) model. We attribute: a) the redshift 
independence to the heavy element pollutants being disposed into the disc and 
accreted onto the central supermassive black hole (SMBH); b) the limited He excess to 
the accretion-wind metabolism of a top-heavy population of evolving massive main 
sequence stars; c) the super-solar CNO enrichment to the nuclear synthesis during 
their post-main-sequence evolution; d) the large N/O to the byproduct of multiple stellar generations;  and
e) the Mg, Si, and Fe to the ejecta of type II supernovae in the disc.
These results provide supporting evidence for f) ongoing self-regulated star formation, 
g) adequate stellar luminosity to maintain marginal gravitational stability,  
h) prolific production of seeds and i) dense coexistence of 
subsequently-grown residual black hole populations in AGN discs.
\end{abstract}


\section{Introduction}
\label{sec:introduction}

Active Galactic Nuclei (AGNs) are the brightest persistent cosmic beacon in the Universe
\citep{schmidt1963, elvis1994}.  They are thought to be powered by viscous dissipation in accretion 
discs (ADs) around super-massive black holes (SMBHs) \citep{lyndenbell1969}.  They are often 
accompanied by ``star-burst'' phenomena in their host galaxies \citep{sanders1988}.  Sloan Digital 
Sky Survey (SDSS) provides a rich data set to characterize AGNs' luminosity function 
and to infer the evolution of their SMBHs' mass distribution ($M_\bullet \sim 10^{4-9} M_\odot$ 
and $m_{\bullet 8} \equiv M_\bullet/10^8 M_\odot$) 
through the cosmic times from red-shift $z_\gamma\sim 5$ to today \citep{greene2007, Shen2011}.  These models lead to
the determination of the most-probable average energy conversion (from the rest-mass energy to radiation)
efficiency factor $\epsilon_\bullet \sim 0.06$, the ratio of luminosity ($L_\bullet$) to its Eddington 
limit $L_{\rm E \bullet}$ $(\lambda_\bullet \equiv L_\bullet/L_{\rm E \bullet} \sim 0.6)$, the duration 
($\sim 10^8$ yr) of AGN episodes, and the SMBH's accretion rate  
\begin{equation} 
{\dot{M}}_\bullet
\simeq 2 f_{\bullet} m_{\bullet 8}\ {M_\odot {\rm yr}^{-1}}, 
\label{eq:mdotsmbh}
\end{equation}
where the scaling factor
$f_\bullet \equiv (\lambda_\bullet/0.6) (0.06/\epsilon_\bullet)$ \citep{marconi2004, shankar2009}.  

AGNs' spectra contain characteristic broad and narrow emission lines.  It is generally assumed that 
the narrow-line regions (NLR) reside in their host galaxies. Based on the assumption that the 
lines' width reflects the Doppler shift of gas with a random speed comparable to the local Keplerian 
speed $V_{\rm k} (=\sqrt {G M_\bullet/R}$), the broad-line regions 
(BLR) are inferred to be on ADs' surface at a distance $R \gtrsim 10^3 R_\bullet$ where 
$R_\bullet \equiv G M_\bullet/c^2$ is SMBH's gravitational radius.
Detailed models of the lines' equivalent width as well as line ratios provide useful information on the incident ionization 
flux, BLR's density, temperature, and chemical abundance. The reverberation mapping technique based on the long-term monitoring of AGN's multi-wavelength light curves provides a determination of discs' surface temperature 
$T_{\rm e}$ and aspect ratio $h (\equiv H/R$ where $H$ is the thickness at disc radii $R$) 
distribution \citep{blandford1982, fausnaugh2016, starkey2023}.  

The internal structure of the disc is opaque and poorly constrained. The main uncertainty is 
the magnitude of the effective viscosity $\nu$ which regulates angular-momentum-transfer and 
mass-diffusion rates \citep{lyndenbell1969}. The widely adopted ``$\alpha$ model'' is based on 
{\it ad hoc} prescription $\nu = \alpha_\nu H^2 \Omega$ for 
turbulent viscosity where $\Omega=V_{\rm k}/R$ and $V_{\rm k}$ are the Keplerian angular and 
spacial frequency respectively, whereas $\alpha_\nu$ is a dimensionless parameter of order 
unity \citep{shakura1973}.  In a steady state, the mass flux in disk \citep{goodman2003} is given by:
\begin{equation}
    {\dot M}_{\rm d}  = 3 {\sqrt 2} (\alpha_\nu h^3/Q) M_{\bullet} 
\Omega  = {\dot M}_{\bullet}.
\label{eq:mdotdisc}
\end{equation}
where 
$Q \equiv c_{\rm s} \Omega /\pi G \Sigma$ is the gravitational stability parameter
\citep{safronov1960, toomre1964}.

Conventional $\alpha$ disc models led to the realization that regions beyond a few light days
($r_{\rm Q}$ in \S\ref{sec:sepad0} and \S\ref{sec:Appendix A}) contain sufficient mid-plane density $\rho_{\rm c}$ and surface density 
($\Sigma = 2 \rho_{\rm c} H$) to excite instabilities with
$Q  \lesssim 1$ \citep{paczynski1978} and gravito turbulence with $\alpha_\nu \sim 1$ \citep{lin1988}.  
We can also infer from the magnitude of $h
(\lesssim 10^{-2})$ \citep{starkey2023} together with Equations (\ref {eq:mdotsmbh}) 
and (\ref{eq:mdotdisc}) that $Q \lesssim 1$ unless $\alpha_\nu \sim 1$.
This physical condition may lead to persistent star formation \citep{Collin1999, Collin2008, goodman2003, thompson2005,
nayakshin2007, levin2007, lobban2022, Chen2023ApJ}.  These discs and their 
central SMBHs also commonly coexist with dense nuclear clusters of stars \citep{kormendy2013}.
During their passage through the discs, some of these stars may encounter sufficiently
strong drag to sediment onto the discs \citep{syer1991, artymowicz1993}.  The luminosity from the embedded 
stars provides auxiliary energy sources to supplement powers for radiation from the 
disc surface as well as to sustain a state of marginal gravitational stability and to
self-regulate an equilibrium rate of star formation.  Such a scenario has been hypothesized
to account for the origin of S stars in the Galactic Center \citep{davies2020}.  The coincidence
of these stars' age ($\sim 6 \pm 2$ Myr) \citep{ghez2003} and the inferred epoch (a few Myr ago) for the 
onset of the Fermi bubble \citep{su2010} provide a hint that these stars may 
have either formed or were trapped and rejuvenated in a once-active and now-depleted 
accretion disc around the Sgr A$^\ast$ SMBH
\citep{zheng2020, zheng2021}.  

The evolution of these stars also enriches the heavy element content in their host discs.  
An intriguing clue for this process is the discovery of super-solar metallicity inferred 
from the ratio of certain emission lines.  In order to analyze the implication of these
observational data, we introduce and qualitatively outline the basic concept of a Stellar Evolution 
and Pollution in AGN Discs (SEPAD) scenario in \S\ref{sec:sepad0}. 
The main purpose of this paper is to gather and scrutinize available observational
data so they can be used to provide useful constraints on the SEPAD scenario.
We re-examine AGNs' broad line
data to corroborate and constrain the extent of self-chemical enrichment in these discs.  
We first recapitulate, in \S\ref{sec:observations}, conventional methods used in the abundance 
measurement of both $\alpha$ and Fe-peak elements.  We gather 
existing information on the abundance evolution in the BLR and undertake a critical 
revaluation of the available evidence on AGNs' chemical abundances of individual elements
as functions of AGNs' luminosity, red-shift,  and MSBHs' mass. We interpret their 
implication in terms of the SEPAD scenario in 
\S\ref{sec:summary}.  We discuss the observational clues on a) the 
most likely location of metallicity production, b) the embedded-star population in the accretion 
disc, and c) the depository of relic heavy elements around AGNs. 
We outline some uncertainties, suggest additional investigations and
tests for the SEPAD model.  


\section{Basic concept of the SEPAD model}
\label{sec:sepad0}
In this section, we briefly describe the conceptual framework and present some 
expressions for relevant disc properties based on the SEPAD model. In order to 
avoid distraction from the main theme of this paper, detailed theoretical construction 
and quantitative derivation of these quantities will be presented elsewhere.

Our formulation of the SEPAD model is based on the assumption that 
the inner disc regions close to the SMBH are regulated by
MHD turbulence in accordance with the conventional $\alpha$ model.
Marginal gravitational stability (with $Q 
\simeq h M_\bullet/ M_{\rm d} \sim M_\bullet/2 \pi \rho_{\rm c} R^3 
\simeq 1$ where $M_{\rm d} = \pi \Sigma R^2$ is the characteristic 
mass of the disc gas) is reached at a transitional radius $ r_{\rm Q} 
(\sim 0.06\ \alpha_\nu^{2/9} \kappa^{2/3} m_8 ^{7/9}$~pc where 
$\kappa$ is the opacity in cgs units).

In the outer disc region (at $R \gtrsim r_{\rm Q}$), marginal 
gravitational stability ($Q \sim 1$) is preserved by the energy 
feedback from embedded stars which are either formed {\it in situ} 
or captured in the AGN discs. Under the assumption that angular 
momentum transfer in the disc at $R \gtrsim r_{\rm Q}$ is mainly 
due to gravito-turbulence with $\alpha_\nu \sim 1$ \citep{Lin1987, 
deng2020} we evaluate the radial distribution of $\Sigma$,  
$\rho_{\rm c}$, $c_{\rm s}$, the aspect ratio $h=c_{\rm s}/ V_{\rm k}$,
mid-plane $T_{\rm c}$ and surface temperature 
$T_{\rm e}$ and radiative flux $Q^-$ in AGN discs.  

The deduced aspect ratio $h(R)$ and $Q^- (R)$ are consistent with those inferred from 
the reverberation mapping data \citep{starkey2023}, the microlensing light curve
\citep{cornachione2020} and the observed spectral energy distribution (SED), 
especially over the infrared wavelength range \citep{sanders1989}.  
In discs with $h \sim 10^{-2}$ around $10^8 M_\odot$ 
SMBHs, $M_{\rm d} \sim 10^{-2} M_\bullet \sim 10^6 M_\odot$.
However,
viscous dissipation $Q^+ _\nu$ alone is no longer adequate to balance $Q^- 
(> > Q^+ _\nu)$ as is 
generally assumed in the standard accretion-disc theory.  

The SEPAD model 
assumes that radiation released from the ongoing evolution of embedded stars 
provides an auxiliary source that is needed to maintain a thermal equilibrium 
\citep{goodman2003, thompson2005}.  The stellar energy flux $Q^+ _\star$ is powered by
a) the fusion of hydrogen H into helium He, with a mass-energy conversion efficiency
$\epsilon_{\rm H} = 0.007$, via p-p chain or CNO cycle in 
low-mass or massive main-sequence (MS) stars, b) He into $\alpha$ elements, 
with a mass-energy conversion efficiency
$\epsilon_\alpha = 0.001$, via triple-$\alpha$ or $\alpha$-chain reaction 
during the post-main-sequence 
(PostMS) phase, c) type SN I and SN II supernova explosions, accompanied by 
Fe production, and d) accretion onto remnant stellar-mass black holes (rBHs).  

Since the luminosity $L_\star$ of both MS and PostMS stars are a function of their mass 
$m_\star$, the condition of thermal equilibrium $Q^+ _\star \simeq  Q^- $ 
determines the magnitude of the stars' surface density $s_\star \simeq Q^- / L_\star$ 
for a given stellar mass function.  In contrast to field stars in their host galaxies, 
stars embedded in self-gravitating AGN discs accrete gas at rates
determined by $\rho_{\rm c}$ and $c_{\rm s}$ in the SEPAD model.
As $m_\star$ of MS stars increases, their $L_\star$ approaches 
their Eddington limit ($L_{\rm E \star} = m_\star c^2/\tau_{\rm Sal}$
where $\tau_{\rm Sal} = 4.8 \times 10^8$ yr is the Salpeter time scale) 
and the radiation pressure suppresses their
accretion rate.  These luminous stars produce an intense wind 
which eventually leads to an accretion-wind equilibrium 
with $m_\star \sim {\mathcal O} (10^{2-3} M_\odot)$.  
A population of $N_\star \sim 10^3$ coexisting 
such massive stars (or a larger group of less massive stars) 
are needed to provide adequate $Q^+ _\star$ for balancing $Q^-$ within 
$R=1$pc from a $m_{8} =1$ SMBH. The total mass of this stellar population is a small fraction of $M_\bullet$
and comparable or less than $M_{\rm d}$ of the disc gas.  The corresponding
rate of He production is ${\dot M}_{\rm He} \simeq N_\star m_\star/
(\epsilon_{\rm H} \tau_{\rm Sal}) \sim 0.3 {M}_{\odot} {\rm yr}^{-1}$.


There are some uncertainties in these stars' subsequent 
evolution. If their wind returns their He-byproduct to the disc 
and they accrete fresh H replenishment to their nuclear burning zone,
one generation of these massive stars may remain indefinitely 
on the MS and continually pollute the disc with only He byproducts
\citep{cantiello2021} with a change $\Delta Y_{\rm d} \sim 
{\dot M}_{\rm He}/{\dot M}_{\rm d} \sim 0.15$ add to the 
primordial $Y_0 = 0.245$ after the big bang \citep{peimbert2016} and elevate the disc
He fraction (by mass) to $Y_{\rm d}=Y_0 + \Delta Y_{\rm d} \sim
0.4$.  These massive MS stars do not produce extra 
$\alpha$ or Fe, and they do not evolve into rBHs.

But, if the stellar wind is confined in the
stars' proximity and efficiently re-accreted, 
their internal He fraction and molecular weight $\mu_\star$ 
would increase. Moreover, if the accretion and growth of embedded
stars can be quenched \citep{dobbsdixon2007, li2021}
by gap formation in the disc \citep{lin1993}
with $m_\star \lesssim 480$, they would have an
outer radiative zone, which would provide a mixing
barrier between the mass-exchange region and nuclear
burning interior and cut off replenishment to the
latter.  
The accretion-wind equilibrium is maintained 
with a decreasing mass until they evolve off the MS track 
with a mass $m_\star \sim {\mathcal O} (30)$ 
(Ali-Deb, Cummings, and Lin, in preparation).   
PostMS stars convert He into $\alpha$ elements 
at a rate ${\dot M}_\alpha$ and emit a luminosity 
$L_{\rm PostMS} \simeq \epsilon_\alpha {\dot M}_\alpha c^2 
\sim L^\star _{\rm E}$ near their Eddington limit.
Similar to the MS evolution, an accretion-wind
equilibrium is maintained with a decreasing mass
and a mass-loss rate ${\dot M}_{\rm PostMS} \sim {\dot M}_\alpha$
as both $Z_\alpha$ and $\mu_\star$ increase.
On a time scale $\tau_{\rm PostMS} \simeq M_{\rm PostMS} 
/ {\dot M}_{\rm PostMS} \lesssim {\mathcal O} (1$ Myr), 
wind loss from stars
reduces their $M_{\rm PostMS}$ to ${\mathcal O} (15-30)
M_\odot$ just before core-collapse. 
The total mass of $\alpha$ elements produced in their 
cores and released to the disc is $\Delta M_\alpha \sim {\mathcal O} 
({\rm \ a \ few} M_\odot)$ \citep{cantiello2021}. The conversion 
to Fe does not significantly reduce the net $\Delta 
M_\alpha$ stars release before core-collapse.
Type II SNe produce Fe-rich ejecta and rBHs with 
masses $\Delta M_{\rm Fe} \sim m_\bullet
\sim {\mathcal O} ( 0.1 M_\odot)$ 
\citep{sukhbold2016, rodriguez2021}.  

Self-regulated 
formation of next-generation stars follows the demise 
of these stars over a (MS+PostMS) lifespan $\tau_\star \sim$ a few Myr.  
Over AGN's active phase $\tau_\bullet \sim 10^8$ yr, 
the He, $\alpha$, and Fe pollutants are carried by the 
gas, with some turbulent diffusion, inwardly flowing
through the disc and accreted by the central SMBH.  Due to 
the He-to-$\alpha$ and $\alpha$-Fe conversion as well
rBHs formation, 
\begin{equation}
\Delta Y_{\rm d} \sim 
{\dot M}_{\rm He}/{\dot M}_{\rm d} - \Delta Z_\alpha
- \Delta Z_{\rm Fe} - \Delta_\bullet
\label{eq:deltaacount}
\end{equation} 
where $\Delta Z_\alpha \simeq N_\star \Delta M_\alpha / 
{\dot M}_{\rm d} \tau_\star \sim {\mathcal O} ((2-3) Z_{\alpha \odot})$,
$\Delta Z_{\rm Fe} \simeq N_\star \Delta M_{\rm Fe} / {\dot M}_{\rm d} 
\tau_\star \sim {\mathcal O} (\rm Z_{\rm Fe \odot})$, 
the mass conversion into residual black holes per each 
generation of stars is
$\Delta_\bullet \simeq N_\star M_{\rm rBHs} / {\dot M}_{\rm d} 
\tau_\star$ where the initial mass of the rBHs is a fraction
of the pre-collapse cores' mass.  
These yields lead to $Y_{\rm d}=Y_0 + \Delta Y_{\rm d}$ slightly 
larger than the solar value, super-solar $Z_\alpha \sim 
\Delta Z_\alpha $, and nearly solar  $Z_{\rm Fe} \sim \Delta Z_{\rm Fe} $.

The triple-$\alpha$ and $\alpha$-chain reactions also lead 
sub-solar $\rm [N/(C+O)] \lesssim -1$.  But this ratio is enhanced 
through the CNO burning in the next generation stars.  
Repeated core-collapse events lead to $\Delta_\bullet \sim \Delta Z_\alpha$
per generation and steadily increases embedded-rBHs population
throughout the AGN active phase.  Although they do not modify the 
disc composition, their accretion of
disc gas contributes to the auxiliary energy budget  
and their coalescence leads to intense bursts of
gravitational waves.

A comparison between the metallicities of AGN discs extrapolated from 
their broad emission lines and the divergent compositional inferences
based on these two stellar-evolution scenarios can potentially be used to 
identify the most likely SEPAD pathway.  The compositional evolution, 
if any, can also be used to characterize the disc flow and verify
the {\it in situ} contamination hypothesis. Another prediction of the
SEPAD model is that the narrow-line regions NLR (in the outermost 
disc region or in the host galaxies) is less metal-rich than the
broad-line regions BLR (which contain the embedded stars).  
The production, diffusion, and advection of the heavy-element 
pollutants reach an equilibrium as they are carried by the inflowing 
gas and deposited in the SMBH.  The saturated metallicity along
this assembly line does not accumulate and is independent of $z_\gamma$.

\section{Abundance determinations in AGNs}
\label{sec:observations}

\subsection{Metallicity indicators for the BLR}

The gas emitting the broad lines is the closest material to the central engine that shows a line spectrum 
subject to abundance analysis.  Reverberation mapping gives a BLR radius $\sim 2-300$ light days \citep{bentz2013, Cackett2021, horne2021, Mandal2021}, scaling as the square root of continuum luminosity. The broad lines typically have an FWHM of $\sim5000$~km/s. The kinematic pattern appears to be predominantly orbital motion, combined with some outflow \citep{sturm2018}. The emitting gas is believed to have an electron density $n_e \approx 10^{10}~ \rm cm^{-3}$, inferred from photoionization analysis and from the absence of broad forbidden lines \citep{davidson1979}. A consequence of this high density is that the mass in the BLR emitting gas is quite small, only a few $\sim 100$ solar masses.  This is a tiny fraction of the mass of the accretion disc or the presumed mass of galactic stars present at similar radii in the nucleus.  The BLR gas is widely assumed to come from the surface of the accretion disc, either as a photoionized atmosphere or a radiation-driven wind (e.g., \cite{Czerny2019}, and references therein).  Therefore, chemical abundances in the BLR are the best available proxy for the composition of the disc itself.

In their original discussion of star trapping in AGN discs, \cite{artymowicz1993}
cited indications of super-solar and red-shift-independent BLR abundances as 
support for disc enrichment by trapped stars.   However, abundances in the BLR 
are difficult to measure.  The forbidden lines are suppressed by high density, 
and ultraviolet lines of elements such as C, N, O, Mg, Si, and Fe must be used.  
The upper levels of these lines have high excitation potentials, and their 
intensity relative to the recombination lines is 
sensitive to temperature.  There are no good line ratios available to 
measure the electron temperature, on account of the faintness of the upper-level 
transitions, the large width of the lines, and the complicating effects of high 
density and optical depth. Chemical abundance measurements thus rely on the judicious use 
of suitable line ratios, together with photoionization modeling that takes 
account of the thermal balance and ionization structure of the emitting gas.

Early work on AGN abundances often used nitrogen as a secondary indicator of 
the overall metallicity in the BLR, largely relying on the \nv~$\lambda1240$ line. 
\cite{hamann1993} analyzed published line ratios out to red-shifts $z_\gamma 
\sim 2$ with the aid of galactic chemical evolution models, see also \cite{hamann2002}. 
They found a metallicity of $Z_{\alpha}\sim 1-10~Z_{\odot}$, the higher values being 
associated with higher redshift and luminosity.  They inferred that chemical 
evolution in massive galactic nuclei proceeded rapidly in the early universe.  
Less clear was whether this enrichment came from normal stellar populations 
or from processes intrinsic to the AGN.

Since the work of \cite{hamann2002}, the subject of abundance in AGN has 
received increasing observational and theoretical attention. A key development 
is the use of primary indicators of the abundance of $\alpha$-elements in the 
BLR, without relying on nitrogen as a secondary indicator.  This advancement 
is significant because there are serious discrepancies between abundances 
inferred from the \nv\ line and other indicators, as discussed below.  A promising 
example is the intensity ratio of \nagao\, which \cite{Nagao2006a} 
found to vary approximately as $Z_{\alpha}^{0.4}$ on the basis of photoionization 
models computed with the Cloudy program \citep{Ferland2017}.  This ratio 
has been used in a number of subsequent studies and will be used here.  
See Appendix \ref{sec:Appendix A} for a discussion of the photoionization physics 
underlying this indicator.

\subsection{The \(\alpha\)-element abundance}
\label{sec:alphablr}

Numerous studies have found that the $\alpha$-element abundance in the BLR is high and does not evolve with cosmic time. \cite{Nagao2006a} measured UV line ratios from SDSS spectra and compared with photoionization models.  They found a typical metallicity of $\sim 5 Z_{\odot}$, with little dependence on redshift in the range $z_\gamma 
= 2.0 - 4.5 $, but a factor two increase in metal abundance with increasing luminosity.  Although \cite{Nagao2006a} included line ratios involving nitrogen, these conclusions may be inferred from the \nagao\ ratio alone. \cite{Jiang2007} obtained spectra of 5 quasars at $z_\gamma =5.8 - 6.3$ and found similar values of \nagao\ and \nvciv\ to the results of \cite{Nagao2006a}. \cite{Xu2018} analyzed composite SDSS spectra binned by $z_\gamma$, $M_\bullet$, and $L_\bullet$, using \nvciv\ and \nagao\ as abundance indicators. They found supersolar metallicity, systematically increasing with $M_\bullet$ and $L_\bullet$ but no redshift dependence. \cite{Yang2021} and \cite{Wang2022} present spectra of overlapping samples of quasars at $z_\gamma \sim 6$, finding little evidence for redshift evolution in comparison with published work at lower redshift.

Recent studies have given attention to the profile of the \civ\ line as an indicator of outflows that complicate the interpretation. \cite{Temple2021} analyzed the ultraviolet line ratios of a large sample of SDSS quasars at 
$z_\gamma=2 - 3.5$.   They found that the blueshift of \civ\, reflecting the strength of the blue wing relative to the symmetrical core, increases with the Eddington factor $\lambda_\bullet$ (see also \cite{Temple2023}).  Correlated with this trend is an increase in \nagao. \cite{Temple2021} suggested that the blue wing is emitted by dense gas in a wind originating close to the black hole, and that only the core of the line profile should be used for abundance analysis. \cite{Lai2022} present spectra of 25 quasars at $z_\gamma \sim 6$, 
also finding no redshift evolution.  For those with \civ\ blueshift less than $1500 \rm \, km\,s^{-1}$, \cite{Lai2022} found a mean value \nagao\ = 0.26, corresponding to $Z_{\alpha} \approx 3.3 Z_{\odot}$ according to the modeling by \cite{Nagao2006a}. Larger values, $\sim 10 Z_{\odot}$, are inferred for the objects with higher \civ\ blueshift, if the flux in the entire line profile is used.  The \nvciv\ ratio implies still higher values, even for the objects with low \civ\ blueshift, on the assumption that nitrogen behaves as a secondary element.

The issue of the \civ\ profile also bears on the question of correlations between the BLR $\alpha$-element abundance and $M_\bullet$ or $L_\bullet$ \citep{Temple2021, Lai2022}. \cite{Xu2018} reported a strong correlation of both $M_\bullet$ and $L_\bullet$ with the BLR metallicity inferred from \nagao\ and \nvciv\ (see also \cite{Nagao2006a}).  However, \cite{Temple2021} argue that these trends vanish in an analysis that allows for the blue wing of \civ. Figure 3 in \cite{Temple2021} shows the line intensity ratios \nvciv\ and \nagao\ in the composite spectra. For fixed \civ\ blueshift velocity, the line intensity ratios do not have a distinguishable correlation with $M_\bullet$, $L_\bullet$, or $\lambda_\bullet$. This conclusion is also supported by \cite{Lai2022}.

\begin{figure}
\centering
\includegraphics[width=0.47\textwidth]{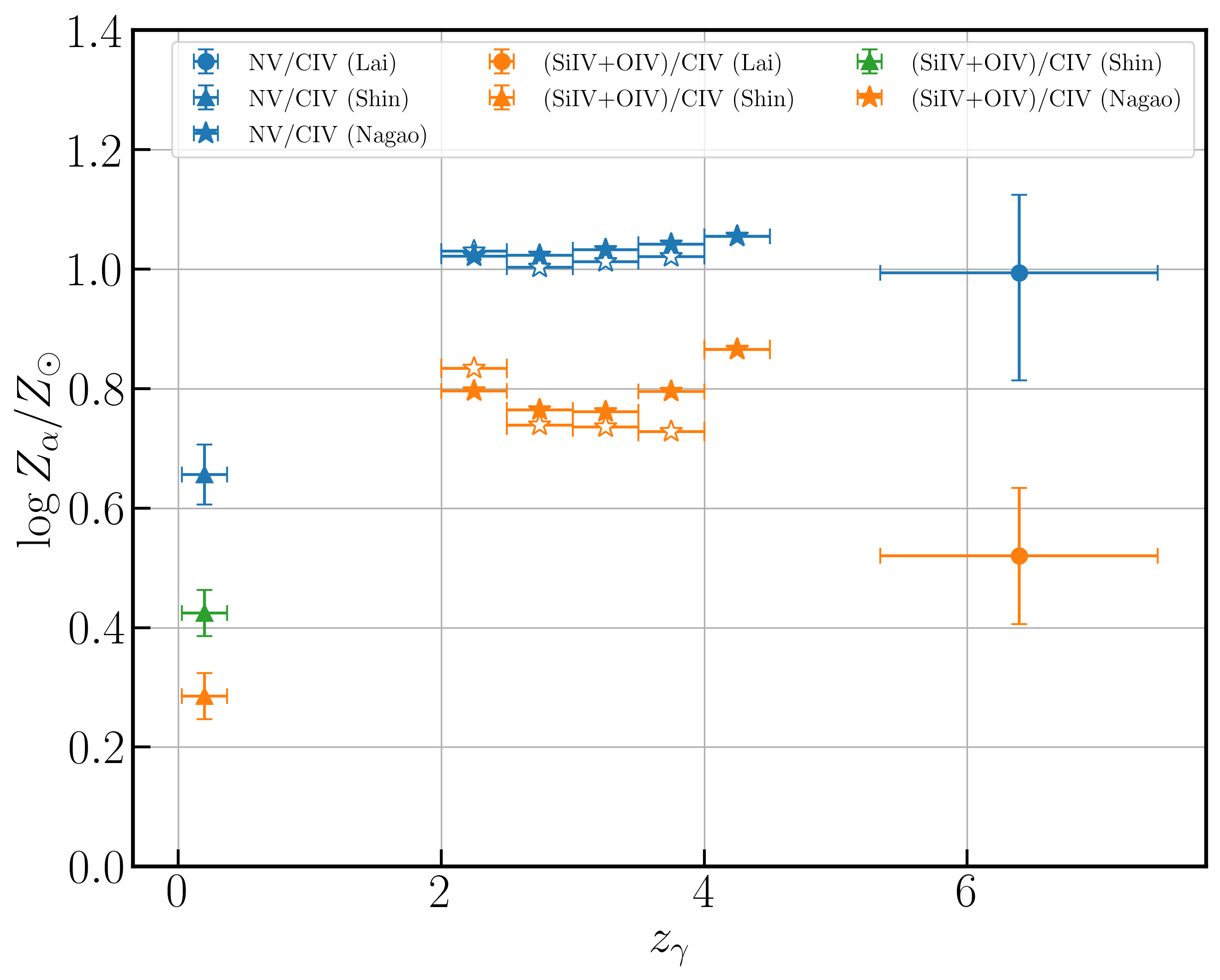}
\caption{Inferred $\alpha$-element abundance as a function of redshift, $z_{\gamma}$. The different redshift bins are from the measurements in \protect\cite{Shin2017} (low-$z_{\gamma}$, triangles), \protect\cite{Nagao2006a} (intermediate-$z_{\gamma}$, stars), and \protect\cite{Lai2022} (high-$z_{\gamma}$, circles).
The filled blue and orange data points correspond to the mean value alpha abundance inferred from \nagao\ and \nvciv\ (with the assumption of secondary nitrogen production), and the error bars are the standard deviation of the scatter. The open stars correspond to the objects in \protect\cite{Nagao2006a} with intermediate luminosity ($-25.5 > M_B \geq -28.5$). Since the line intensity ratio \nagao\ does not depend linearly on the $\alpha$-element abundance, the mean of the inferred abundance for the individual object is not the same as the inferred abundance from the mean of the intensity line ratio. For a better comparison, we added inferred $\alpha$-element abundance from the mean value of \nagao\ as the green triangle point.}
\label{redshift_evolution}
\end{figure}

Results for $Z_\alpha$ are illustrated in Figure \ref{redshift_evolution}, where several reported abundances for the $\alpha$-elements are plotted as a function of redshift.  The value $Z_\alpha = 3.3 Z_{\odot} $ at $z_\gamma \approx 6$ represents the results of \cite{Lai2022} derived from \nagao\ for objects with a small blueshift of \civ. (Specifically, we give the average of  the two values of $Z_\alpha$ derived by \cite{Lai2022} from their composite spectra for their lowest two bins in \civ\ blueshift.)  The point at low redshift is based on \cite{Shin2017}, where we have used the model results of \cite{Nagao2006a} to convert \nagao\ to $Z_\alpha$, object by object, for the 17 objects for which this line ratio is given by \cite{Shin2017}. The plotted value is the mean of the 17 values, and the error bar is the standard error of the mean derived from the scatter among the objects.  (Uncertainties in the calibration of the line ratio to abundance are not included.)  None of the 17 objects has a \civ\ blueshift greater that $1500 \rm \,km\,s^{-1}$. A comparison of the low and high redshift points with small \civ\ blueshift indicates a limit on evolution in $Z_\alpha$ from redshift 0.2 to redshift 6 of less than $\pm 0.2$~dex.  However, there are further indications of minimal evolution with redshift.  At intermediate redshift, Figure \ref{redshift_evolution} shows results for composite spectra for several redshift bins from \cite{Nagao2006a}. These values are all more metal-rich than the high and low redshift points, but the \cite{Nagao2006a} results do not distinguish objects with high and low \civ\ blueshift. Importantly, among themselves, the \cite{Nagao2006a} points show imperceptible redshift evolution. Figure 25 of \cite{Nagao2006a} indicates a spread of only about 0.05~dex in \nagao\ as a function of redshift at fixed luminosity, and even this small scatter shows no systematic trend with redshift.  This suggests a limit of 0.2~dex evolution in $Z_\alpha$ over the redshift range $2.25 < z_\gamma  < 4.25$. The results of \cite{Nagao2006a} show a similarly small scatter in \nvciv\ among the same redshift bins, also without a systematic trend with redshift (at fixed luminosity).  Although there is a troubling discrepancy between values of $Z_\alpha$ derived from  \nvciv\ versus \nagao\ (see below), \nvciv\ may still be useful as an indicator of the redshift evolution.  The absence of any redshift trend for \nvciv\ implies a limit of $\sim 0.1$~dex on any trend of $Z_\alpha$ over this redshift range, based on the strong dependence of  \nvciv\ on $Z_\alpha$ for secondary nitrogen as assumed by \cite{Nagao2006a}.  The results of \cite{Xu2018} further reinforce the impression of negligible evolution in $Z_\alpha$  (less than 0.1~dex) over the redshift range $2.5 < z_\gamma  < 5$. 

There is found to be no correlation between the \nvciv\ and \nagao\ ratios with the far-infrared (FIR) luminosity of AGN, which indicates that BLR $\alpha$-element abundance does not correlate with the star forming rate in the host galaxy \citep{2010MNRAS.407.1826S}. This may be another indication that the $\alpha$-element is produced locally from star formation near the black hole, rather than from the host galaxy.

Because of discrepancies between \nvciv\ and other nitrogen line ratios (e.g., \niv$\lambda 1486$/\civ\ and \niii $\lambda 1750$/\oiiis $\lambda 1664$ \citep{hamann2002}, also discussed below in \S \ref{N_abundance}), we adopt the ratio \nagao\ as our preferred abundance indicator for the $\alpha$-elements. In summary, for objects with small \civ\ blueshift, this gives $Z_\alpha \approx 3 Z_{\odot}$, with a variation from high to low redshift of no more than 0.2~dex. 

\subsection{Nitrogen Abundance in the BLR}
\label{eq:nitrogenblr}
\subsubsection{Photoionization model with simplified LOC model}
\label{4-loc}
We use the locally optimally emitting clouds (LOC) to model the BLR clouds \citep{Baldwin1995, Korista2004}. The LOC model assumes a distribution of densities of BLR clouds, which have a range of distance from the central continuum source. With its range of density and ionizing flux, the LOC model potentially mitigates the sensitivity of the emission lines to physical conditions.  We have examined the predicted intensity of the emission lines as a function of abundance in a simplified LOC model consisting of four components with $n_{\rm H} = 10^{10}$ or $10^{11}\rm cm^{-3}$ (for the text below, we use $\rm cm^{-3}$ as the unit for particle densities, such as $n_{\rm H}$, $n_{\rm e}$, etc.) and incident ionizing flux $\phi = 10^{19}$ or $10^{20} \rm photons~cm^{-2}~s^{-1}$ (for the text below, we use $\rm photons~cm^{-2}~s^{-1}$ as the unit for $\phi$). The individual models were computed with Cloudy 17.03 using a slab geometry, the ``strong bump'' continuum of \cite{Nagao2006a}, a turbulence of $ v_{\rm turb} = 10^7 $ cm s$^{-1}$ and stopping column density of $N_{\rm H} = 10^{23}$ cm$^{-2}$ \citep{Baldwin1995}, and iteration of the diffuse radiation. The four models were weighted to give equal contributions to the H$\beta$ flux in the combined LOC model. 

\subsubsection{Inferred Nitrogen abundance}

\label{N_abundance}
Previous work on QSO abundances has largely involved the \nv~$\lambda1240$ line.  However, a number of workers have noted a systematic discrepancy between abundances derived from \nv\ in comparison with other line indicators such as \nagao\ and those involving \niv\ and \niii\ (e.g., \cite{hamann2002, Xu2018}).  In a study of NLS1 galaxies, where \nv\ is cleanly separated from Ly$\alpha$, \cite{Shemmer2002} found that the N/C ratio derived from \niv\ was systematically lower by a factor of 3 or 4, and they suggested that \nv\ may not be a reliable abundance indicator.  We focus on the \niii\ line for nitrogen abundance.

As discussed by \cite{shields1976} and others, ratios of the UV collisionally 
excited lines to each other are less sensitive to the electron temperature 
than are the ratios of these lines to the recombination lines of H and He.  
Therefore, a photoionization model with an approximately correct overall 
metallicity should provide an adequate estimate of the electron temperature 
for analyzing the appropriate line ratios.  At the same time, the 
models provide ionization fractions for the ions involved.  This allows us 
to estimate the relative abundance of the heavy elements simply by scaling 
selected line ratios of the model to match those observed.  The \niiioiii\ 
ratio was found to be reliable by \cite{hamann2002} on the basis of a grid 
of photoionization models. These lines are weak, and reliable measurements 
are few.  \cite{dietrich2000} report \niii\ and \oiiis\ intensities for 
four QSOs at $z_\gamma \sim3$.  These have an average of \niiioiii\ 
= 0.72, with a standard error of the mean of 0.10~dex.  Our LOC model 
with $3Z_{\odot}$ and solar N/O gives \niiioiii\  = 0.50; and scaling 
from this, the implied value of [N/O] = 0.16 (or (N/O)/(N/O)$_\odot$ = 1.44).  (Scaling from our reference Cloudy model gives a closely similar result.)  The 
results of \cite{wills1995} give a similar average value of \niiioiii\ 
= 0.66. The composite NLS1 spectrum of \cite{constantin2003} gives 
\niiioiii\ = 0.49, which implies [N/O] = -0.009. 
A more careful calculation by changing the nitrogen abundance for the four-component Cloudy LOC model with $3Z_{\odot}$ (see \S \ref{4-loc}) gives [N/O] = 0.13 (or (N/O)/(N/O)$_\odot$ = 1.35) for \niiioiii\ = 0.72. This result does not differ much from the scaling method, because the \niiioiii\ ratio scales, to first order, linearly as a function of [N/O], as changing the nitrogen abundance does not have a large effect on the electron temperature.

These 
modest values stand in contrast to the high values derived from N~V 
intensities in many studies, and they suggest that nitrogen falls 
short of full secondary (quadratic) behavior as a function of $Z_\alpha$.  
Note, however, that [N/O] = 0.16 
with $Z =3Z_{\odot}$ means 
[N/H] = 0.60
, so that substantial enrichment in nitrogen over 
the solar value has occurred.

How does this result compare with the implications of the other nitrogen lines? As noted above, the \nv\ line tends to give much larger values of N/C. In contrast, the \niv\ line at $\lambda 1486$ tends to give a low value of N/C.  Let us consider the ratio \nivciv, which involves ions occupying similar volumes in the ionization structure. Tables $3 - 7$ of \cite{Nagao2006a, Nagao2006b} give an average observed value of  \nivciv\ = 0.023, with a dispersion of a factor 2 among the various composite spectra.  The composite spectra of \cite{boyle1990} and \cite{constantin2003}
give similar values.  For comparison, our LOC model (3$Z_{\odot}$ with solar N/C) gives \nivciv= 0.076, and the reference model gives 0.070.  Using the LOC results in the scaling procedure described above, we find (N/C)/(N/C)$_\odot$ = 0.30.  A more careful calculation by changing the nitrogen abundance for the four-component LOC model with $3Z_{\odot}$ (described above) gives [N/C] = -0.47 (or (N/C)/(N/C)$_\odot$ = 0.34). This is smaller than the value of N/O implied by \niiioiii, and much smaller than N/C values typically found from the \nv\ line.  Thus, as previous studies have noted, the three nitrogen lines give very different results.  However, some of this discrepancy may involve the C/O abundance ratio, which we examine below.

The number of reliable measurements of \niii\ is small, and therefore we turn to the \nv\ line to assess the evolution of nitrogen with redshift.  
As discussed above, there is some question about the reliability of abundances derived from \nv\ in an absolute sense.  However, it is quite possible that \nvciv\ still varies roughly linearly with N/C in a relative sense.  Making this assumption, we see that there is little systematic difference between the median value of \nvciv\ in the low redshift sample of \cite{Shin2017} and the high redshift sample of \cite{Lai2022}.  Our restricted luminosity subset of the \cite{Nagao2006a} composite spectra align well with the \cite{Shin2017} and \cite{Lai2022} samples, and show no systematic redshift trend among themselves.  




\subsection{Oxygen Abundance in the BLR}
\label{eq:oxygenblr}
Oxygen is represented in QSOs by two fairly weak emission multiplets,  \oiiis~$\lambda1663$ and \oiv~$\lambda1402$.  The latter is typically blended with \siiv, as discussed above.  In some objects, the \ovi~$\lambda1035$ line is also observed, but this line is especially sensitive to ionization conditions in the BLR.   For the four objects from \cite{dietrich2000} discussed above, the average value of  \oiiiciii\ is 0.28.  Our LOC model gives a value of 0.67, and the scaling procedure then gives (O/C)/(O/C)$_\odot$ = 0.41. Scaling using our reference model gives a similar result. The ratio \oiiiciv, by the same procedure, gives a closely similar result, reflecting the fact that the model value of \ciiiciv\ is close to the observed value. 

How reliable is the inference of sub-solar O/C?   The ratio \oiiiciii\ varies considerably among individual objects.  However, a number of studies give values of \oiiiciii\ or \oiiiciv\, which, averaged over a set of objects, or measured from composite spectra, are similar to the value used above 
\citep{constantin2003, boyle1990, Dietrich2003}.
Could the model overestimate the fractional ionization of $\rm \opp$ and thereby 
exaggerate the line intensity, so that a lower oxygen abundance seems necessary?  
Our reference Cloudy model, which approximates the observed \ciiiciv\ ratio, 
gives volume-averaged ionization fractions of $\rm \left \langle X(\cpp) \right \rangle / \left \langle X(\hep) \right \rangle =  0.81$ and  $\rm \left \langle X(\opp) \right \rangle / \left \langle X(\hep) \right \rangle =  0.96$. This is a large fraction of the volume of the $\rm \hep$ zone where these ions normally exist, and there is not much room to increase the volume of either ion.  If a large part of the $\rm \opp$ were converted to $\rm \oppp$, the resulting intensity of \oiv\ would greatly exceed the observed intensity of the \oiv\ + \siiv\ blend.  However, as discussed in many studies (e.g., \cite{hamann2002, shuder1979}), \ciii\ is sensitive to collision de-excitation at densities above $10^9$.  Our reference model has $n_{\rm H} = 10^{10}$, giving substantial collisional de-excitation of \ciii.  The critical density for the \oiiis\ ($n_{\rm cr} = 3.13\times 10^{10}$ at $T_e=10^4~\rm K$, see Table 1 in \cite{hamann2002}) ion is considerably larger, and it will be less affected at the model density.  If the typical density in the BLR were less than $10^{10}$, this would give a lower value of \oiiiciii\ that would seem to imply a low O/C abundance ratio.  Thus, we cannot make a definitive case for O/C less than solar.  However, most modern BLR models assume a density of $10^{10}$ or higher, guided in part by reverberation mapping results for the BLR radius (see discussion in \cite{Sameshima2017}).  Moreover, the low O/C value is indicated by our LOC model, which spans a range of densities and ionizing fluxes; and the LOC model has had success in accounting for AGN spectra (\cite{Baldwin1995, Korista2000}).  The possibility of a non-solar O/C ratio therefore merits further study.

If O/C does indeed have the subsolar value estimated here, this would partly account for the lower value of N/C compared with N/O found above from  \nivciv\ and \niiioiii.

\subsection{BLR Helium Abundance}
\label{sec:blrhe}

Helium has a number of emission lines in the optical, infrared, and ultraviolet.  These result largely from radiative recombination but can be affected by collisional excitation and radiative transfer \citep{Osterbrock2006}. For AGN, this necessitates comprehensive photoionization models along with constraints on $n_e$ and $\phi$.  For typical BLR parameters and AGN ionizing continua, the $\rm \hepp$ ion is confined to an inner fraction of the volume of ionized gas, and $\rm \hep$ occupies the bulk of the emitting volume.  The most useful line is \hei$~\lambda$5876, which is fairly strong and falls in an uncrowded part of the spectrum. However, this generally requires a low redshift, giving limited opportunities to measure abundances of He and heavier elements in the same object.

For a metallicity of $Z_{\alpha} = Z_{\odot}$ or $Z_{\alpha} = 3 Z_{\odot}$, we ran the quartet of models with a helium abundance by number of atoms of He/H = 0.10, 0.15, 0.20, and 0.40. Table \ref{tab:helium_intensity1a} gives the resulting intensity of \hei~$\lambda5876$ relative to H$\beta$ for the combined LOC model.  The \hei~$\lambda5876$ intensity is not hugely affected by the metallicity, and it varies fairly strongly with helium abundance ($\propto$(He/H)$^{0.7}$).  This supports the use of \hei~$\lambda5876$ as an abundance indicator. 

\begin{table}
\begin{center}
\caption{Four-component LOC results (described in \S \ref{4-loc}) of \hei~$\lambda5876$/H$\beta$ intensity ratio with varying helium abundance, for $Z_{\alpha} = Z_{\odot}$ and $Z_{\alpha} =  3Z_{\odot}$.}
\label{tab:helium_intensity1a}
\begin{tabular}{l c r}
\hline
$Z/Z_{\odot}$ & He/H
& $I$(\hei$~\lambda$5876)/$I$(H$\beta$)
\\
\hline \hline
1 & 0.10 & 0.128
\\
1 & 0.15 & 0.180
\\ 
1 & 0.20 & 0.229
\\
1 & 0.40 & 0.378
\\
\hline
3 & 0.10 & 0.127
\\
3 & 0.15 & 0.178
\\
3 & 0.20 & 0.227
\\
3 & 0.40 & 0.370
\\
\hline
\end{tabular}
\end{center}
\end{table}

\begin{table}
\begin{center}

\caption{Four-component LOC (see \ref{4-loc}) results for \hei~$\lambda$5876, H$\beta$ intensities (in $\rm erg\,cm^{-2}\,s^{-1}$), and line intensity ratio $I$(\hei$~\lambda$5876)/$I$(H$\beta$) for $Z_{\alpha} = Z_{\odot}$, $\rm He/H=0.1$ by number.
}
\label{tab:helium_loc}

\begin{tabular}{l r r r r}
\hline
$\log n_{\rm H}$ & $\log \phi$
& $\log I($\hei$~\lambda$5876) & $\log I($H$\beta$) & ratio
\\
\hline \hline
10 & 19 & 5.984 & 6.973 & 0.102
\\
10 & 20 & 6.982 & 7.876 & 0.128
\\ 
11 & 19 & 6.184 & 7.094 & 0.123
\\
11 & 20 & 7.205 & 8.009 & 0.157
\\
\hline
\end{tabular}
\end{center}
\end{table}

\begin{figure}
\centering
\includegraphics[width=0.47\textwidth]{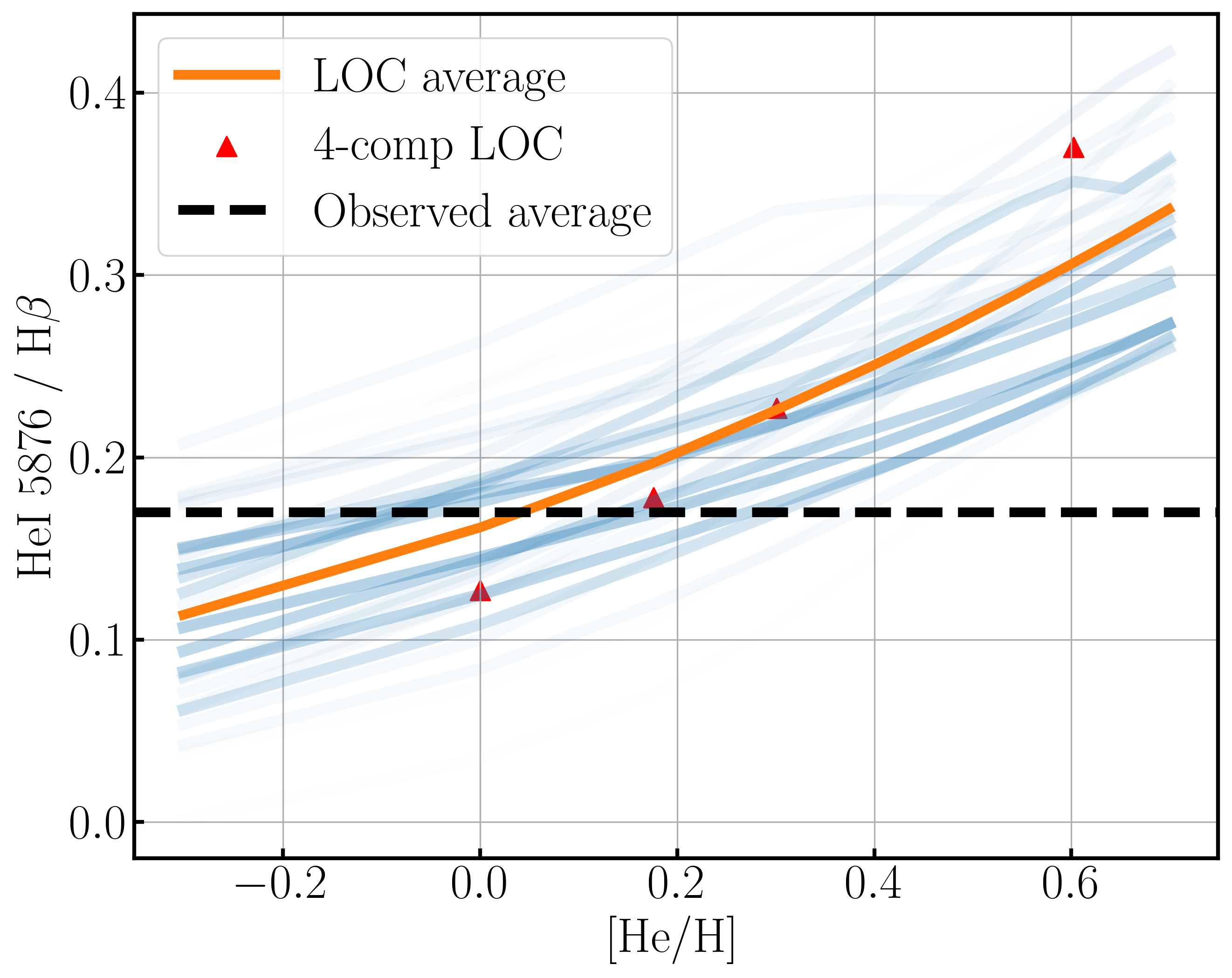}
\caption{Cloudy results of \hei~$\lambda$5876/H$\beta$ for different He/H but at fixed $Z=3Z_{\odot}$. The orange curve shows the weighted average for the line intensity ratios for individual one-zone models (in blue), ranging from $19< \log{\phi} <21$ and $9< \log{n_{\rm H}}<11$, where the weight is given by a Gaussian distribution centered at $\log n_{\rm H}=10$, $\log \phi=19.5$, with a standard deviation $\sigma = 0.5~\rm dex$ in both $n_{\rm H}$ and $\phi$. The shade of the curves corresponds to the weight of the average. The red triangles are the result of a 4-component LOC model with $Z_{\alpha} = 3Z_{\odot}$, as shown in Table \ref{tab:helium_intensity1a}.}
\label{LOC_avge_He}
\end{figure}
We investigated the helium abundance using the quasar sample from the Sloan Digital Sky Survey (SDSS) DR16 \citep{Lyke2020}. We used quasars at redshifts $z_\gamma < 0.4$ to ensure coverage of both \hei~$\lambda$5876 and H$\beta$, with additional criteria of high median signal-to-noise ratio (S/N) across all good pixels in a spectrum ({\texttt{SN\_MEDIAN\_ALL} $> 15$}) \citep{Lyke2020}, plus the line width (FWHM) of H$\beta$ larger than $500~\rm km/s$, the selected sample contains a total number of 6053 quasar spectra. Then we used {\ttfamily PyQSOFit} \citep{PyQSOFit_Guo} to decompose the spectra into quasar and host-galaxy components, then fit the underlying continuum, \hei~$\lambda$5876, H$\beta$, and \oiii~$\lambda$4959, 5007 for the quasar component. Multiple Gaussian functions were used for each line. The lines are fitted with a narrow (NLR) and a broad (BLR) component for the \hei~$\lambda$5876 and H$\beta$. For the broad component, we used two Gaussian functions corresponding to the very broad component and the emission from the intermediate line Region (ILR) \citep{Hu2008, Adhikari2016}. For \oiii~$\lambda$4959, 5007, four Gaussian functions were used to fit the core and wing of the line profile. Due to the presence of the broad Na~{\sc i}~$\lambda$5890, 5896 emission, \hei~$\lambda$5876 lines can be contaminated \citep{Thompson1991}, so we estimate the uncertainty to be as much as 30\%.
Our measurements of AGN from SDSS show that the observed \hei~$\lambda5876$/H$\beta$ ratios in low redshift AGN generally fall in the range of 0.10 to 0.20, with an average of about 0.17 and an occasional value as high as 0.30, 
which agree with published studies, 
including \citep{VandenBerk2001, Shang2007, Bentz2010, 
Kollatschny2018}.  Figure \ref{LOC_avge_He} shows the LOC average (in orange) and the one-zone models (in blue) for the \hei~$\lambda5876$/H$\beta$ intensity ratio for different helium abundance. The LOC results therefore suggest that He/H in typical AGN is around 0.13 but may range as high as 0.30 (see also Table \ref{tab:helium_intensity1a}).

\begin{figure}
\centering
\includegraphics[width=0.47\textwidth]{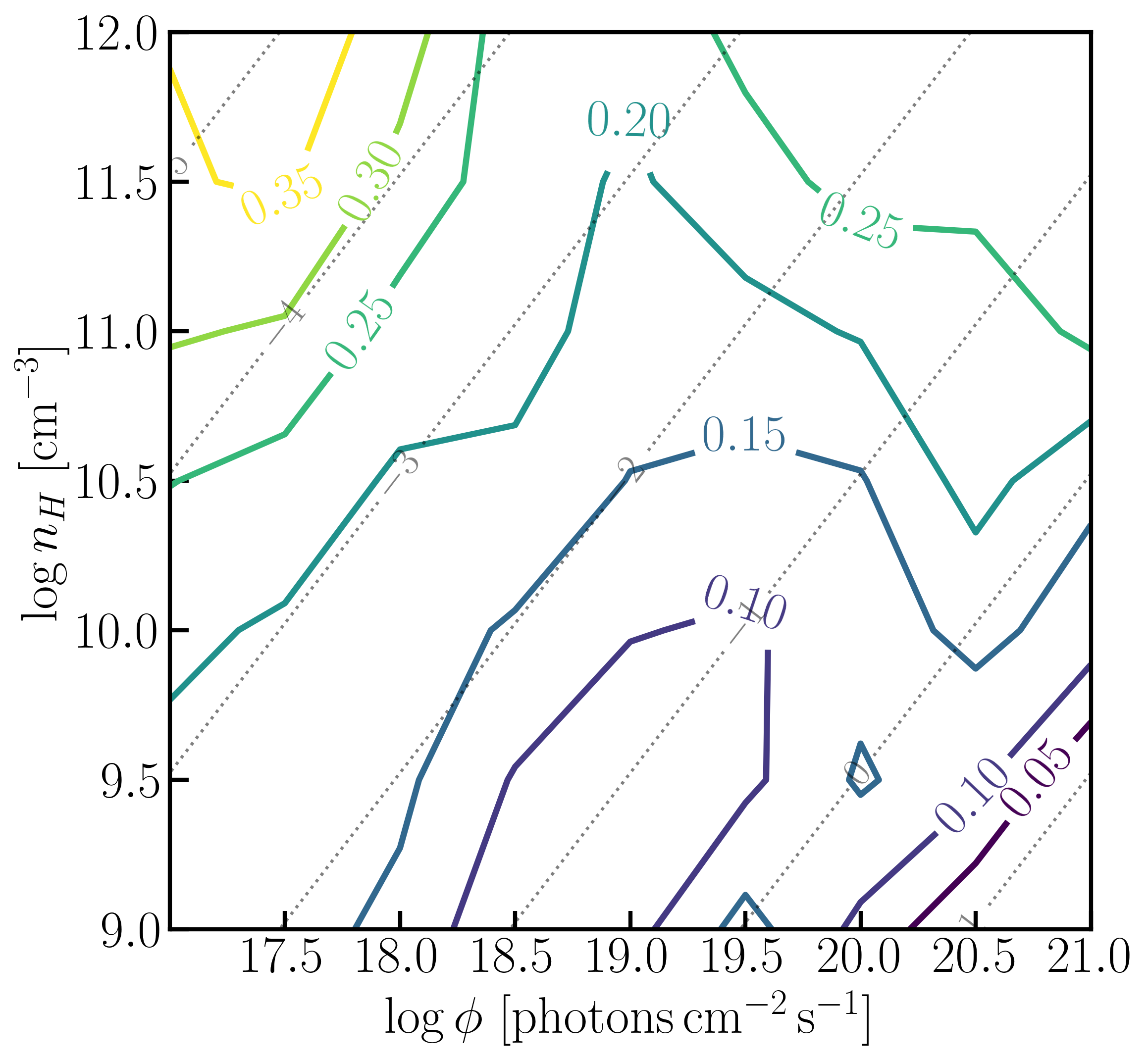}
\caption{Cloudy model for \hei~$\lambda5876$/H$\beta$ intensity ratio with stopping column density $\log N_{\rm H} = 23~\rm cm^{-2}$, $Z_\alpha = 3Z_{\odot}$, and $\rm He/H=0.1$ by number. The ionizing flux varies from $17< \log{\phi} <21$ and density varies from $9< \log{n_{\rm H}}<12$. The grey dotted lines show the ionization parameter, $\log U$. For a given ionizing flux, the HeI 5876/H$\beta$ intensity ratio increases for higher density.}
\label{Contour_he1_hb}
\end{figure}

\begin{figure}
\centering
\includegraphics[width=0.47\textwidth]{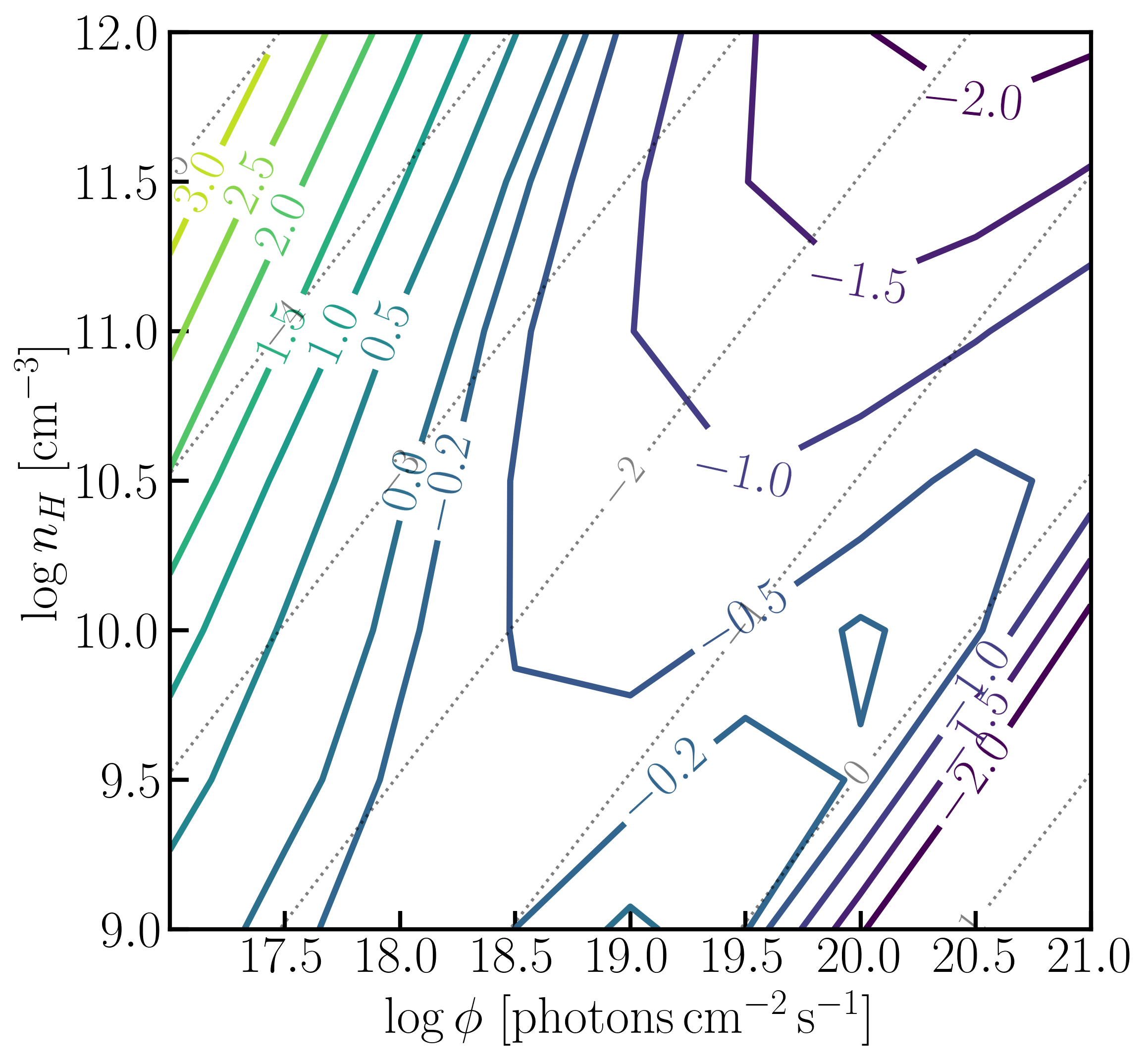}
\caption{Same as Figure \ref{Contour_he1_hb}, but for $\log[ I($\ciii~$\lambda1909$) / $I($\civ~$\lambda1549$)]. The observed value is around -0.5 to -0.2.}
\label{Contour_C3_C4}
\end{figure}

How reliable are these inferences?  Table \ref{tab:helium_loc} gives the \hei~$\lambda5876$ intensity for the four component models of the simple LOC model.  The \hei~$\lambda5876$/H$\beta$ ratio for solar metals and He/H = 0.10 
ranges from 0.102 for ($\log n_{\rm H}$, $\log \phi$) = (10, 19) to 0.157 for (11, 20).  Higher density 
in particular tends to give stronger \hei~$\lambda5876$. Could much of the BLR involve gas under conditions that enhance \hei~$\lambda 5876$?  Figure \ref{Contour_he1_hb} shows the \hei~$\lambda5876$/H$\beta$ 
ratio for one-component Cloudy models over a range of $n_{\rm H}$ and $\phi$, computed as described 
above. Here, we adopt a stopping column density of $N_{\rm H}=10^{23}~\rm cm^{-2}$. But for models with high ionization parameters ($\log U \gtrsim -0.5$), the choice of stopping column density can affect the \hei~$\lambda5876$/H$\beta$ ratio, due to the truncation of H+ Strömgren sphere at $N_{\rm H}=10^{23}~\rm cm^{-2}$. For example, the model ($\log n_{\rm H}$, $\log \phi$) = (10, 20) with solar metals and He/H = 0.10 gives \hei~$\lambda5876$/H$\beta=0.128$ when $N_{\rm H}=10^{23}~\rm cm^{-2}$, but the same model with $N_{\rm H}=10^{24}~\rm cm^{-2}$ gives \hei~$\lambda5876$/H$\beta=0.103$. In addition, for even higher ionization parameters ($\log U \gtrsim 0$), both H+ and He+ Strömgren sphere are not complete at $N_{\rm H}=10^{23}~\rm cm^{-2}$, resulting in a low \hei~$\lambda5876$/H$\beta$ ratio (as shown in the bottom-right corner of Figure \ref{Contour_he1_hb}). The LOC average (see Figure \ref{LOC_avge_He}), on the other hand, is not hugely affected by the choice of $N_{\rm H}$, as it spans a range of ionization parameters. The contours in Figure \ref{Contour_he1_hb} also show that, for a helium abundance by number of He/H = 0.10, the intensity ratio ranges from 
0.05 to 0.35, encompassing the full range of observed values.  However, the \hei\ emission 
comes from the bulk of the ionized volume. Any one-zone model for the \hei\ emission 
must give realistic intensities for other lines that largely arise in the same zone,
including \ciii\ and \civ.  Figure \ref{Contour_C3_C4} shows log~ \ciii~1909/\civ~1549 
for the same models as Figure \ref{Contour_he1_hb}.  Observed values typically fall in 
the range of -0.2 to -0.5. The figure shows that values in this range occur in a wedge-shaped 
region  region, centered around $\log n_{\rm H}=9.6$ and $\log\phi = 18.3$, along with 
two narrow arms extending to high $n_{\rm H}$ and $\phi$. However, these upper arms are 
not realistic regarding other features of the BLR spectrum.  In the upper left arm, \ciii\ 
is weak relative to H$\beta$ due to collisional de-excitation at high density, and \civ\ 
is weak due to a low ionization parameter $U$.  This region also overproduces low 
ionization lines such as \mgii. Such considerations eliminate the parameter space giving the highest and lowest values of \hei~$\lambda5876$/H$\beta$. 

We may also ask whether the AGN with the strongest He I lines have any indication of unusual physical conditions. \cite{Shang2007} give line intensities in the rest frame optical and ultraviolet for 22 broad line AGN.  The data show no significant trend in the \ciiiciv\ or the \mghb\ ratio as a function of \hei~$\lambda5876$/H$\beta$, which argues against a systematic difference in physical conditions for objects with especially strong He I emission.

The above considerations argue against high density as the cause of the occasional large \hei~$\lambda5876$/H$\beta$ values, and we therefore entertain the possibility of a high helium abundance.  The strongest observed values of \hei~$\lambda5876$/H$\beta$ are 0.30 or slightly higher. The LOC results quoted above give \hei~$\lambda5876$/H$\beta$ = 0.30 for He/H = 0.27. Conventional galactic chemical evolution cannot account for such a high abundance of helium. If we adopt a metallicity of $3 Z_{\odot}$ (see above), then a helium abundance He/H = 0.11 would be implied by the primordial value He/H = 0.083 together with a Galactic enrichment ratio (expressed in terms of the mass fraction) of $\rm \Delta Y/\Delta Z = 1.75$ (see eq. \ref{eq:deltaydeltaz}).  Most of the helium enhancement in the highest helium AGN must therefore come from some process that can produce a large amount of helium with a relatively small yield in heavier elements.  We discuss below the prospect that SEPAD may provide such a process.

\subsection{The Fe/Mg ratio}
\label{sec:feovermg}

The abundance of iron in the BLR has received much attention, not least because the highest 
redshift quasars approach a time when Fe production by SN~Ia is problematic (see \S\ref{sec:galacticha}-A1 and A2). 
The \feii\ emission, especially in the optical, differs dramatically from object to object. Theoretical efforts to explain the intensity of \feii\ involve a complex model of the $\rm \fep$ ions with many energy levels and transitions.  The work of \cite{Wills1985} showed that fluorescence between different transitions has an important effect, along with collisional excitation and continuum fluorescence. The recent work of \cite{sarkar2021}, using the latest \feii\ atomic data sets in Cloudy, largely succeeds in reproducing the optical and UV multiplet intensities observed in AGN.

Studies of the iron abundance in the BLR and its redshift dependence have largely focused on the ultraviolet \femg\ ratio.  
The \mgii~$\lambda 2800$ line and the UV \feii\ bands around 2400~\AA\ are close in wavelength 
and involve similar stages of ionization. These studies generally find that \femg\ depends little, if at all, on redshift \citep{Maiolino2003, Verner2009, DeRosa2011, Schindler2020, Yang2021, Wang2022}.  However, caution is warranted for several reasons. In the recent photoionization models by \cite{sarkar2021}, 
as Fe/Mg varies from 0.1 to 10 times solar, the \feii\ (UV)/\mgii\ ratio only increases by $\sim$0.4 dex, i.e., 
\begin{equation}
I(\text{\feii})/I(\text{\mgii}) \propto (\text{Fe/Mg})^{0.19}. 
\label{sarkar_Fe_over_Mg_ratio}
\end{equation}
Subtle trends in the line ratio could imply substantial trends in abundance. The emission in \feii(UV) comes from 
deeper in the photoionized clouds than does \mgii\ (e.g., Fig. 7 of \cite{Sameshima2017}), 
making the ratio potentially sensitive to details of the BLR cloud structure. The \femg\ line ratio varies 
with Eddington ratio, approximately as $\lambda_{\bullet}^{0.3}$ \citep{Sameshima2017, Shin2021}.
Other parameters, such as microturbulence, also affect the \femg\ ratio \citep{sarkar2021}.
Furthermore, the sensitivity of the \mgii\ line to Mg/H is greater than the sensitivity of \feii\ to Fe/H; in our one-slab Cloudy models, \femg\ varies as (Fe/Mg)$^{0.7}$ when only Mg/H is varied. Thus, the \femg\ ratio cannot be 
interpreted in isolation.

Another issue is the choice of template used to fit the \feii\ blends 
in the spectrum.  The empirical template by \cite{VW01} (VW01) is set 
to zero under the \mgii\ line, where \feii\ could not be measured.  Several recent authors 
have used the template by \cite{T06} (T06), which uses theoretical spectra to 
fill in the \feii\ emission under \mgii. This reduces the \mgii\ intensity and increases \femg.  \cite{Schindler2020} give a detailed discussion.  All of the measurements discussed here use the T06 template or make some other allowance for \feii\ emission under the \mgii\ line.

\begin{table*}
\centering
\label{tab:Fe_Mg_ratio}
\caption{
The mean value of \femg\ ratio across redshift bins. The second column shows the mean observed value, and the third column shows the adjusted value by applying Equation (12) in \protect\cite{Sameshima2017} with each object's Eddington ratio. For the low redshift objects in \protect\cite{Shin2021}, we also listed the mean \femg\ for four high-Eddington ratio objects ($\log \lambda_{\bullet}>0.5$) to make a better comparison for the high redshift objects.
}

\begin{tabular}{l c c c c}
\hline
redshift
& (\femg)$_{\rm obs}$
& (\femg)$_{\rm adj}$ & No. objects & reference
\\
\hline \hline
$z_{\gamma} < 0.367$ & 2.88 (3.80) & 4.17 (3.63) & 13 (4)
& \cite{Shin2021}
\\
$0.7<z_{\gamma}<2.3$ & 3.13 & 3.60 & 69,202
& \cite{Shen2011}
\\ 
$4.5<z_{\gamma}<4.7$ & 4.62 &- & 5
& \cite{Dietrich2003}
\\
$4.5<z_{\gamma}<5.1$ & 2.95 & 2.84 & 10
& \cite{DeRosa2011}
\\
$5.7<z_{\gamma}<6.4$ & 4.56 & 4.37 & 28
& \cite{Wang2022}
\\
$5.8<z_{\gamma}<6.4$ & 3.03 & 2.44 & 14
& \cite{DeRosa2011}
\\
$5.9<z_{\gamma}<6.9$ & 5.47 & 5.11 & 32
& \cite{Schindler2020}
\\
\hline

\end{tabular}
\label{tab_iron_mg_ratio}
\end{table*}

In Table \ref{tab_iron_mg_ratio}, we show averages of published measurements of \femg\ in 
several redshift bins.  We have adjusted the observed values of \femg\ to a common fiducial 
Eddington ratio of $\log\lambda_{\bullet} = -0.55$, using a slope $\Delta \log$ (\femg) $= 0.3~\Delta \log 
\lambda_{\bullet}$ from \cite{Sameshima2017}.  This adjustment, which was applied to individual quasars 
before averaging over redshift bins, is particularly important for the low redshift samples, 
\cite{Shin2021} give results for 29 low redshift AGN ($z_{\gamma}
 < 0.367$) with archival 
HST spectra.  Of these, the thirteen with $\log\lambda_{\bullet} > -1.5$ have an average
\femg\ = (2.88, 4.17) respectively (before, after) correction to the common Eddington ratio. 
For the brightest four objects, with $\log\lambda_{\bullet} > -0.5$, the corresponding values are (3.80, 3.63). 
For a moderate redshift sample, we used the database of spectral measurements 
by \cite{Shen2011} for SDSS DR7 quasars (see also \cite{Shin2021}).  
For 69,202 quasars at $0.7 < z_{\gamma} < 2.3$, the mean \femg\ is (3.13, 3.60). 
At higher redshift, \cite{Dietrich2003} give an average \femg\ of 4.62. (We have applied the $10\%$ adjustment estimated by \cite{Dietrich2003} to allow for \feii\ emission under the \mgii\ line. These values are not adjusted to our fiducial Eddington ratio because the authors do not quote values of $\lambda_{\bullet}$, but the correction is likely small for these bright quasars.)  We have divided the measurements by \cite{DeRosa2011} into two redshift bins.  The average of the three high redshift studies is \femg\ = 4.2, after adjustment for $\lambda_{\bullet}$.  This agrees within $\sim 0.1$~dex with the results at lower redshift.  Given the dependence of \feii(UV)/\mgii\ as $(\rm Fe/Mg)^{0.2}$, the corresponding limit on any redshift dependence of Fe/Mg is about $\pm0.5$~dex.

What is the actual value of Fe/Mg that corresponds to this redshift-independent line ratio? Let us consider the highest 3 redshift bins in Table \ref{tab_iron_mg_ratio}, with an average \femg\ = 4.2.  For this value, Figure 11 of \cite{sarkar2021} gives Fe/Mg of 0.27 times solar (for the \cite{DeRosa2011} integration range).  That figure, however, is for solar abundance (except for iron).  For our adopted alpha-element abundance of three times solar, adjustments must be made.  The Cloudy models underlying Figure 11 of \cite{sarkar2021} use log $\phi$ = 20 and log $n_{\rm H}$ = 11, giving $\log U = -1.5$.  In our Cloudy models, these parameters give \mghb\ in agreement with typical observed values $\sim1.7$ (based on the composite spectrum of \cite{VandenBerk2001} and the sample of SDSS quasars described above.) 
However, when all heavy elements, including iron, are increased together, \mghb\ increases and \femg\ decreases. In order to restore agreement for \mghb, the ionization parameter must be increased. In our Cloudy models, a reduction in density to $\log n_{\rm H} = 10.5$, giving $\log U = -1.0$, achieves this.  These parameters raise \femg\ by about 0.05 dex, and then Equation (\ref{sarkar_Fe_over_Mg_ratio}) requires a reduction of Fe/Mg by 0.25 dex to restore agreement for \femg. The result of this analysis is [Fe/Mg] $= -0.8 \pm 0.5$, using the above uncertainty range.  For an alpha-element abundance of three times solar, we then have [Fe/H] $= -0.3 \pm 0.5$.  This rules out a Pop II value of Fe/H, but it is compatible with a Pop II (i.e., SN II) value of Fe/Mg.

This result, however, is based on a single photoionization model with a particular $n_{\rm H}$ 
and $\phi$.  How might the result change for a mix of parameters as envisioned by the LOC model? 
Suppose that the mix contains some highly ionized gas that emits much of the radiation 
in \civ, \ovi, etc., as well as much of the observed H$\beta$.  Then the bulk of the \mgii\ 
and \feii\ must come from a component with a high value of \mghb, so that the mix gives the 
observed ratio. As an example, consider a low-ionization component that has $\log U = -1.5$ 
as discussed above but with $Z_{\alpha} = 3Z_{\odot}$.  For these parameters, in 
our Cloudy models,  \femg\ is now 0.13~dex lower (at fixed Fe/Mg) than for solar metals. 
By equation (\ref{sarkar_Fe_over_Mg_ratio}), we now must raise the above value Fe/Mg = 0.27 
from Figure 11 of \cite{sarkar2021} by 0.65~dex to fit the observed value of \femg, 
resulting in Fe/Mg = 1.2(Fe/Mg)$_{\odot}$.  Thus, simply having \mgii\ and \feii\ come 
from a region of modestly lower ionization parameter leads to a much larger value of 
Fe/Mg.  This suggests that the values of Fe/Mg and Fe/H derived in the preceding paragraph 
should be considered lower limits, and it illustrates the need for caution in deriving an 
iron abundance from the UV \feii\ emission.
\begin{figure}
\centering
\includegraphics[width=0.47\textwidth]{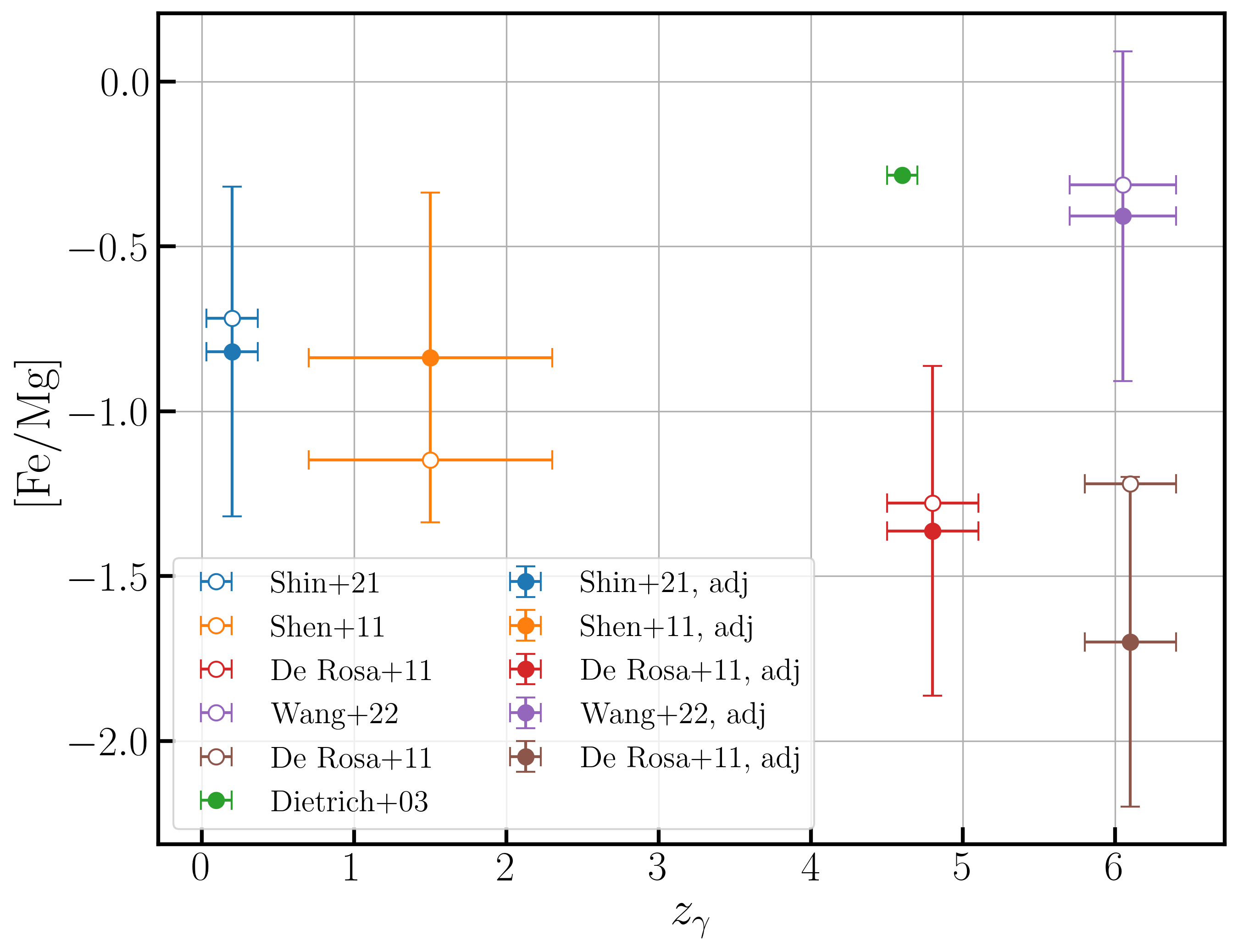}

\caption{
Redshift evolution of the BLR iron abundance. The open circles correspond to the inferred iron abundances [Fe/Mg] using Figure 11 in \protect\cite{sarkar2021}. The closed circles are the inferred iron abundances using the adjusted line ratio \femg\ with the Eddington correction in \protect\cite{Sameshima2017}. \protect\cite{Dietrich2003} does not have data for the Eddington ratio, so we use the un-calibrated \femg\ to calculate [Fe/Mg].
}
\label{redshift_evolution_Fe}
\end{figure}

\subsection{Comparison of NLR and BLR}
\label{sec:blrnlralpha}

The region emitting the narrow emission lines of AGN (Narrow Line Region or ``NLR'') affords an opportunity for comparison of BLR abundances with interstellar abundances in the core of the host galaxy but outside the accretion disc.  The NLR has a radius of order 100~pc for moderate luminosity AGN and scales roughly as $L_{\rm AGN}^{1/2}$ \citep{bennert2006a}. (This refers to the radius containing the bulk of the emission, not the outermost boundary.) For a density $n_e = 10^4~\rm cm^{-3}$ (e.g., \cite{terao2022}), the mass of emitting gas is of order $10^4$ or $10^5$ $M_{\odot}$.  This is much larger than the mass of emitting gas in the BLR but less than the mass of the accretion disc or the stellar mass located at the radius of the NLR (see \S\ref{sec:sepad0}).

Studies of abundances in the NLR have increased in recent years, but uncertainties remain.  The NLR has typical gas densities ranging from $10^3$ to $10^6~ \rm cm^{-3}$. While these densities are much lower than in the BLR, they still exceed the critical densities of many of the forbidden lines observed in the optical and neighboring wavelength bands.  This makes it difficult to derive electron temperatures from the forbidden line ratios often used in nebular studies, and this in turns impedes direct determinations of abundances. Consequently, photoionization models are needed. Because the narrow lines are difficult to measure in broad line objects, especially in the ultraviolet, such studies have often targeted narrow line objects (Seyfert 2 or QSO 2).  The results are assumed to apply to the NLR of broad line objects, in a statistical sense, on the basis of the unified model of AGNs.  

\cite{Nagao2006b} carried out a photoionization model study of narrow line AGN in the redshift range $z_{\gamma} = 1.2$ to 3.8, using rest frame ultraviolet lines, in particular \heii, \ciii, and \civ.  They found that the observations were consistent with a low density ($10^3~\rm cm^{-3}$) NLR with subsolar abundances (0.2 - 1.0 $Z_\odot$ or a high density model with results ranging from subsolar to supersolar (0.2 - 5 $Z_\odot$).  In either case, there was no indication of evolution with redshift (see also \cite{Matsuoka2009}).
\cite{Mignoli2019} studied a large sample of Type 2 AGN at redshift 1.5 to 3.0 with the aid of photoionization models.  The results, averaged over three redshift bins, show a progressive decrease with increasing redshift:  12 + log O/H = 8.55, 8.34, 8.16  at $z_{\gamma}$ = 1.7, 2.2, 2.9, respectively.  In contrast, \cite{terao2022} analyze rest-UV spectra of 15 narrow line radio galaxies at redshift 3, including a number of fainter emission lines that help to constrain photoionization models.  They find NLR abundances in the range $Z_{\alpha}$ = 0.6 to 2.1 $Z_\odot$, with an average of 1.2 $Z_\odot$. \cite{Dors2020} report solar or modestly subsolar oxygen abundance for low redshift SDSS narrow line AGN, derived from a variety of semi-empirical calibrations of the strong lines (\nii\ and \oiii). Aside from individual outliers, these studies do not find NLR alpha abundances similar to the high values found for the BLR. In addition, \cite{Du2014} report a positive correlation between \nvciv\ in the BLR and [N {\sc ii}]/H$\alpha$ in the NLR, implying a strong correlation between NLR and BLR metallicities. This correlation can be explained by the metal-rich gas in the BLR being transported to the NLR due to outflow \cite{Du2014}.

The abundance of iron in the NLR can be derived from the intensity of the emission-line ratio of [Fe~{\sc vii]}~$\lambda6087$ to [Ne~{\sc v}]~$\lambda3425$ \citep{Nussbaumer1970}.  This line ratio was measured by \cite{shields2010} in composite spectra of low-redshift SDSS quasars ($z_\gamma \approx 0.3$) grouped into 5 bins by increasing strength of the broad \feii\ emission.  Their results give an average value of \feviinev\ = 0.34, with a scatter of only 10 \% among the bins and no discernible trend with broad \feii\ strength. We ran plane-parallel Cloudy models for the NLR with solar abundances, $\log n_{\rm H} = 3$ or $5$ \citep{Nagao2006b}, $\log U = -0.98$ or $-1.98$, and a stopping electron temperature of 3000~K. 
In terms of the choice of the ionization parameters, the observational results for 15 objects in \citep{terao2022} give $-2.0 \lesssim \log U \lesssim -1.0$. \cite{Perez-Montero2019} gives $-2.42 \lesssim \log U \lesssim -1.27$ for the NLR. In addition, reverberation mapping result \citep{Peterson2013} shows that the majority of the emission from NLR comes from its inner region ($R_{\rm NLR}\sim 1-3 \rm pc$). This first and hitherto only measurement provides the estimated NLR size, independent of those inferred from spatially resolved imaging and spectroscopy \citep{Cackett2021}. This indicates that, within the same AGN, the ionizing flux NLR sees is $\sim 10^6$ times smaller than in the BLR. Combining with $\log n_{\rm H} = 5$, we get $\log U \sim -1.98$. 
The four models gave iron abundance ranging from (Fe/Ne) = 0.85 to 1.55 (Fe/Ne)$_\odot$.
We conclude that the Fe-to-$\alpha$ ratio in the NLR of typical low redshift quasars is close to the solar value. This result and the average line ratio used here are consistent with \cite{Nagao2003}. From their photoionization modeling, we estimate an uncertainty of $\pm0.2$~dex in the derived Fe/Ne ratio.  Our value of Fe/Ne is consistent with our derived value of Fe/Mg in the BLR, within the large error bar of the latter.

An interesting case is that of the ``nitrogen loud'' AGNs, which have exceptionally strong lines of \nv, \niv, and \niii\ from the BLR, suggestive of very high metallicity and secondary nitrogen production. These objects are only a small fraction of the AGN population (e.g., \cite{Batra&Baldwin2014, Maiolino2023}). 
\cite{Matsuoka2017} use the equivalent width of \oiii\ to argue that N-loud AGN have approximately solar metallicity in their NLR (see also \cite{Araki2012}).  

In summary, results for the NLR indicate subsolar or around solar abundances, with a possible decrease toward higher redshift.  There is little indication in the NLR of abundances resembling the high ($\ge 4Z_\odot$) values found for the BLR. Furthermore, the extreme nitrogen abundances shown by the broad lines of some objects appear not to be reflected in their NLR, at least regarding the alpha-elements.  These results support the idea that the extreme metal enrichment of the BLR has its origin in the immediate vicinity of the accretion disc.  

\begin{figure}
\centering
\includegraphics[width=0.47\textwidth]{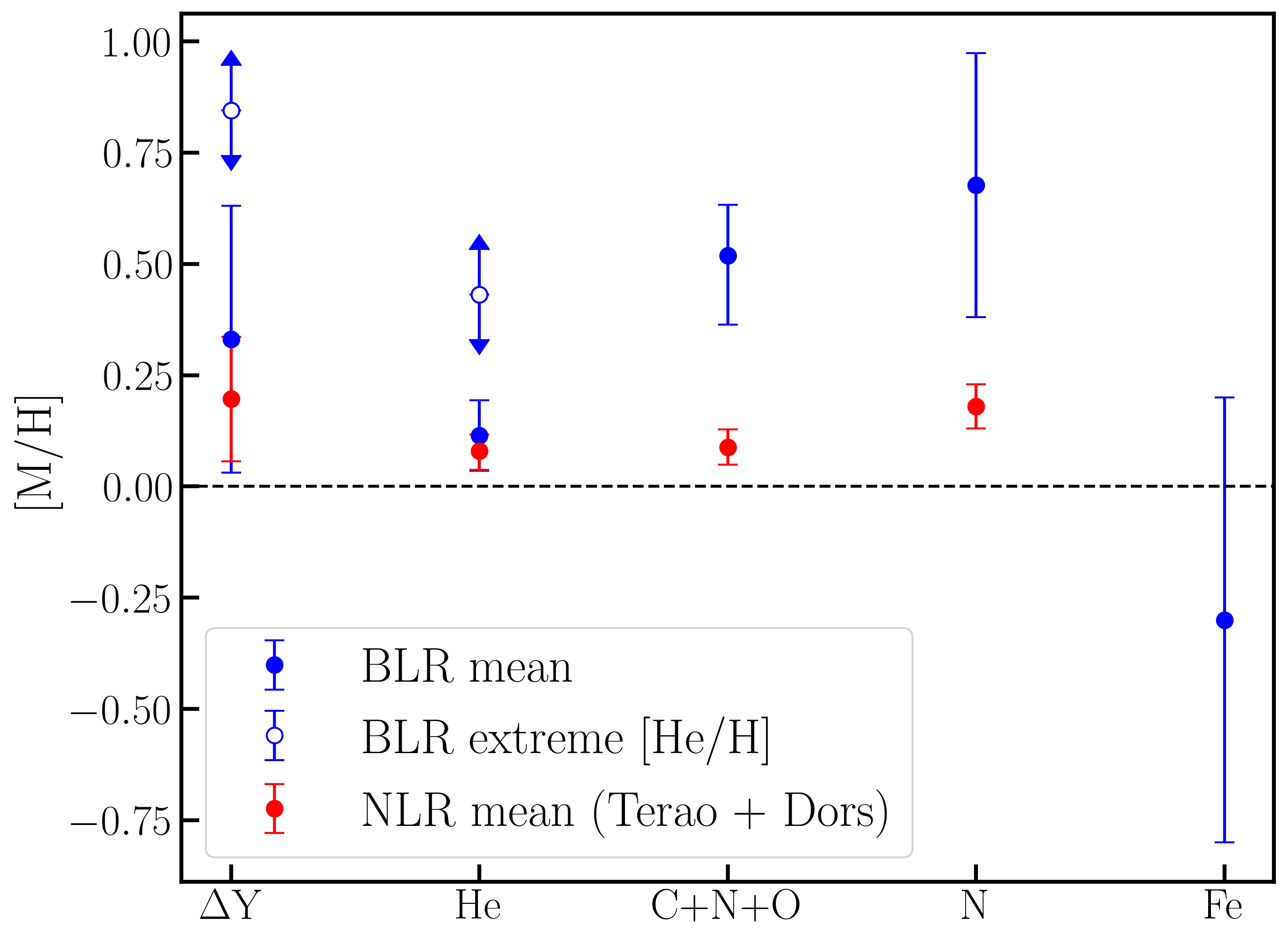}
\caption{Summary of abundance by elements in the BLR (in blue) and NLR (in red). $\Delta Y$ is the enrichment in the AGN helium abundance relative to the primordial helium abundance, normalized by the enrichment in the solar helium abundance, $\Delta Y_G$. The open blue circles represent the inferred helium abundance in the objects with extreme \hei~$\lambda5876$/H$\beta$ values (e.g., 3C120), and the error bars for the extreme Helium abundance corresponds to the modeling error, so the actual error is larger than the error indicated in the plot. The value of [(C+N+O)/H] in the BLR is inferred from the average line intensity of \nagao\ in \protect\cite{Lai2022} with \civ\ blueshift less than $1500 \rm \,km\,s^{-1}$. For the NLR, [(C+N+O)/H] is the average value of the $\alpha$-element abundance in \protect\cite{terao2022}. The value of [N/H] in the BLR is inferred from the line intensity ratio of \niiioiii. For the NLR, [N/H] is inferred from \nii/\oiii\ ratio \citep{Dors2017}.}
\label{abundance_summary}
\end{figure}

\section{Summary and discussions}
\label{sec:summary}

In comparison with the Galactic chemical evolution model (Appendix \ref{sec:galacticha}),  
we interpret their implication of the AGNs' abundance 
in terms of the SEPAD model (\S\ref{sec:sepad0}, \S\ref{sec:sepadimp},
Appendix \ref{sec:galacticha}). 

\subsection{A brief observational synopsis}
\label{sec:synopsis}

The chemical evolution of the Galactic disc and halo is constructed based on the spectroscopy
of populations I and II stars.
The $[\alpha/\rm H]$ and [Fe/H] in these stars correspond to their values in the ISM at their formation epochs, which 
extend over several Gyr (\S\ref{sec:galacticha}).
Similarly, the abundance of the $\alpha$ and 
Fe-peak elements in AGNs with overlapping range of ages ($z_\gamma = 0-7$), is measured from their BLRs 
which are analogous to the ISM in the Galactic context (\S\ref{sec:sepad0}).  

{\it A) $\alpha$-element abundance.}

\noindent
$\bullet$ A1. The most important and robust observational constraint is the lack of evolution 
in BLRs' super-solar [$\alpha$/H] ($\simeq 0.5$ or $Z_\alpha \vert_{\rm BLR} \simeq 0.06 $), 
i.e. its constancy through the cosmic epochs $z_\gamma (\sim 0-7)$ (\S\ref{sec:alphablr}).  

\noindent
$\bullet$A2.  The N/(O+C) ratio is elevated from the solar value.   
N-loud AGNs may represent a small fraction of exceptions (\S\ref{eq:nitrogenblr}).
 
\noindent
$\bullet$A3.  Although some ratios of BLR $\alpha$ lines appear to vary with $M_\bullet$ and
$\lambda_\bullet =L_\bullet/L_{\rm E \bullet}$ \citep{Xu2018}, these dependence can be attributed 
to emission-line physics rather than actual BLR [$\alpha$/H] variations \citep{Temple2021}.  

\noindent 
$\bullet$A4.  As a population, [$\alpha$/H] for the NLR also appears to be $z_\gamma$ independent 
and smaller by $\lesssim 0.5$ (i.e. [$\alpha$/H] $\simeq 0$ or 
$Z_\alpha \vert_{\rm NLR} \sim 0.02$) than that for the BLR (\S\ref{sec:blrnlralpha}).
BLRs' [$\alpha$/H] is also uncorrelated 
with signatures of star formation in the host galaxies \citep{simon10}.

{\it B) Helium abundance.} 
The galactic chemical evolution leads to minor  increases in the Helium mass 
abundance from the big bang $Y_0 \simeq 0.245$ with $Z_{\alpha, 0}=0$ and 
$X_0 \simeq 0.755$ \citep{carigi2008, peimbert2016, valerdi2021}  to today $Y_{\rm G} \simeq 0.28$ \citep{arnett1996} with $\Delta Y_{\rm G} = Y_{\rm G}
- Y_0 \simeq 0.04$ and $\Delta Z_{\rm \alpha, G} = Z_{\rm \alpha, G} \simeq 0.02$ 
so that $\Delta Y_{\rm G} \sim 1.8 \Delta Z_{\rm \alpha, G}$ (Eq. \ref{eq:deltaydeltaz}).

\noindent
$\bullet$  
B1. By number, AGNs' average [He/H] $\sim 0.114$  and in exceptional objects may reach $0.4$ for their BLRs 
(\S\ref{sec:blrhe}, Fig. \ref{abundance_summary}).  The corresponding $Y=0.32$,
$\Delta Y =Y-Y_0 \sim 0.08$ which is slightly smaller than that $1.8 Z_\alpha 
\vert_{\rm BLR} \sim 0.09$ or $\Delta Y_{\rm G} + 1.8 \Delta Z_\alpha \sim 0.1$ extrapolated from 
the BLR's $Z_\alpha \vert_{\rm BLR}$, NLR's $Z_\alpha \vert_{\rm NLR}$, and their 
difference $\Delta Z_{\alpha} (\simeq Z_\alpha \vert_{\rm BLR} - Z_\alpha \vert_{\rm NLR} 
\simeq 0.035)$ with the standard stellar and Galactic chemical evolution models
(Eq. \ref{eq:deltaydeltaz}).  The corresponding  
$[\Delta Y] = {\rm log} (\Delta Y/\Delta Y_{\rm G}) \simeq 0.36$.


\noindent
$\bullet$ 
B2. Within the measurement uncertainties, [He/H] $\sim 0.08 \pm 0.04$ 
by number and $[\Delta Y] \sim 0.25 \pm 0.1$ by mass for the NLR \citep{Dors2022}.

{\it C) Fe-peak abundance.} In the Galaxy, old stars with ([$\alpha$/H] 
$\lesssim -2$) have sub-solar [Fe/Mg] ($\sim -0.3 - 0$) and population I 
stars (with $[\alpha$/H] $\sim 0$) have solar or super-solar [Fe/Mg] ($\sim 0-0.2$).
Since Mg is an $\alpha$ element, this correlation also extends to [Fe/$\alpha$]
and it has been interpreted to indicate that the Fe production was mostly
due to SN II when the Galaxy was young and pristine, whereas the present-day
Fe content is mainly released from SN Ia, which took time to evolve (\S\ref{sec:galacticha}).

\noindent
$\bullet$ 
C1. 
AGNs' slightly sub-solar [Fe/Mg] ($\sim -0.8 - 0.1$) is inferred 
from the broad \femg\ lines (\S\ref{sec:feovermg}).

\noindent
$\bullet$ 
C2. BLRs' nearly-solar [Fe/H] (with a mass fraction $Z_{\rm Fe} \sim 10^{-3}$) 
is inferred from their [Fe/Mg], calibration of 
solar $\alpha$-element-abundance distribution (see above), and super-solar [$\alpha$/H] 
(\S\ref{sec:alphablr}).
 
\noindent
$\bullet$ 
C3. The combination of sub-solar [Fe/Mg] (C1) with super-solar
[$\alpha$/H] (A1) for the BLRs does not match that for the Galactic stars (\S\ref{sec:galacticha}).

\noindent  
$\bullet$ 
C4.  
The inferred [Fe/H] for BLR is independent of $z_\gamma$.

 

D) {\it Observational uncertainties.}

There are some less well established evidences on:

\noindent   
$\bullet$ \
weak, if any, dependence of $\alpha$ element abundance on the SMBHs' mass and AGNs' Eddington ratio (A3) and
 
\noindent   
$\bullet$ \
minor evolution in [Fe/Mg] or [Fe/$\alpha$] element (Fe/Mg) ratio in BLR of low-red shift AGNs.

There is little or no data on:

\noindent   
$\bullet$ \
[He/H] over a wide range of $z_\gamma$ and

\noindent   
$\bullet$ \
[Fe/H] for NLR.

Large observational uncertainties are associated with:

\noindent   
$\bullet$ \
insensitive dependence of \femg\ on [Fe/$\alpha$] and

\noindent   
$\bullet$ \
discrepancies of N/O versus N/C in different lines.

\subsection{Some implications on SEPAD}
\label{sec:sepadimp}
The two salient features of this SEPAD scenario (\S\ref{sec:sepad0}) are: 
1) the prolific production of super solar $\alpha$-element and solar Fe abundances at high-$z_\gamma$ and 
2) the constancy of these abundances since $z_\gamma \sim 7$ (i.e. more than 12 Gyr ago).  
With the observed data (\S\ref{sec:synopsis}), 
we gather supporting evidence and place some constraints on the SEPAD model.

\noindent 
$\bullet$ 
A. {\it Adequate auxiliary power for outer regions of AGN discs.}
The SEPAD model is motivated by the possible auxiliary power, due to thermonuclear 
burning inside the embedded stars, to exceed the energy dissipation rate of the 
accretion flow. 
For a $M_\bullet = 10^8 M_\odot$ SMBH, this transition occurs at a radius ($\sim 3$ light days)  
$\lesssim$ that inferred for the BLR\citep{horne2021, lobban2022}.  Since the polluted BLR is downstream in the accretion-disc flow (i.e. closer to the SMBH) from the sites where He, $\alpha$, 
and Fe elements are being injected into the disc, their stellar sources' 
$Q^+ _\star$ dominates their local $Q^+ _\nu$.  In a thermal equilibrium, 
the supplemental $Q^+ _\star$ leads to extra infrared excess
which is commonly observed in AGNs' SED \citep{sanders1989}.  This inference is supported by 
the half-light radius of AGN discs extrapolated from microlensing observations 
\citep{pooley2007, morgan2018, cornachione2020}, which appears to be larger than that derived based on 
the standard steady-state viscous accretion disc model.  

\noindent 
$\bullet$ 
B. {\it A population of massive MS stars in AGN discs.}
SEPAD models predict a population of MS stars with masses ($m_\star \sim 10^{2-3} M_\odot$)
to maintain an accretion-wind equilibrium. Their nearly Eddington-limited
luminosity is powered by the conversion of H into He at a rate ${\dot M}_{\rm He}$
(\S\ref{sec:sepad0}). The observed super-solar $[\Delta Y] > 0$ in high-$z_\gamma$
($\sim 5-7$) AGNs (\S\ref{sec:synopsis}-B1) supports the expectation of rapid (in 
$\lesssim 1$ Gyr) He enrichment beyond that of normal galactic chemical evolution. 
The mean value of $\Delta Y \lesssim 0.08$ for most AGNs is less than the  
H-to-He conversion rate (${\dot M}_{\rm He}/{\dot M}_{\rm d} \sim 0.15$) required 
to account for the auxiliary power such that some of the He-byproducts during 
the embedded stars' MS evolution may have been converted into
$\alpha$ or Fe elements (\S\ref{sec:sepadimp}-C) 
or be deposited into rBHs (\S\ref{sec:sepadimp}-F).  

\noindent
$\bullet$ 
C. {\it Embedded stars' transition from MS to PostMS evolution.}
An outstanding issue is whether H fuel in massive stars
is continually replenished by a fresh supply of disc gas 
so that they remain on the MS indefinitely or it is 
exhausted in the CNO burning zone, which would lead to 
transition to PostMS evolution (\S\ref{sec:sepad0}).  
Since He is converted into $\alpha$ elements during
the PostMS evolution, the detection of strong signatures 
of super-solar $Z_\alpha$ (\S\ref{sec:synopsis}-A1) 
with weak or no dependence on $\lambda_\bullet$ and $M_\bullet$
(\S\ref{sec:alphablr}, \ref{sec:synopsis}-A3) clearly indicates that this 
transition has taken place 
during the active AGNs' active phase in contrast to the
scenario that massive stars evolve off the MS after
the disc depletion \citep{cantiello2021}.  Moreover, the modestly positive 
observed value of $[Y]$ (\S\ref{sec:synopsis}-B2) also implies that these stars are 
embedded in He-enriched (but not He-predominant) 
disc gas such that their MS-to-PostMS transition cannot
be due to the severe He pollution over wide regions of AGN discs
as suggested by \cite{jermyn2022}.
The quenching of H replenishment requires either an efficient confinement 
and retention of the stellar wind with He-byproducts or  the
prevention of compositional mixing by a radiative zone between the 
recycled gas and the nuclear-burning core (Ali-Deb, Cummings, and Lin
in preparation).  The modest 
$\Delta Y_{\rm d} (\sim 0.08 < {\dot M}_{\rm He} / {\dot M}_{\rm d}\sim 0.15$,
\S\ref{sec:synopsis}-B1)
is consistent with the SEPAD model with evolving (with lifespan 
$\tau_\star \sim 4-5$ Myr rather than ``immortal'') stars 
provided a major fraction of the He-byproduct during massive stars' MS evolution is
retained and recycled (analogous to a close-box model), converted into 
$\alpha$, Fe and released to the disc during their PostMS evolution and subsequent
SN explosions.

\noindent
$\bullet$
D. {\it Mass loss during embedded stars' PostMS evolution.}
The observed $[\alpha$/H] (\S\ref{sec:synopsis}-A1) is comparable to 
the value of $Z_\alpha$ inferred, based on the SEPAD 
scenario, for the $\alpha$-element production during 
the PostMS evolution of $N_\star \gtrsim 10^{3-4}$ coexisting 
embedded stars within a few pc. This agreement
also implies that most of the $\alpha$-byproducts 
is released to the disc along with the He-byproducts
such that the embedded stars had substantial mass lose
along their evolution track and their 
pre-collapse cores have downsized to range 
$\lesssim 15 M_\odot$ before their 
SN II explosion\citep{sukhbold2016}.  

\noindent
$\bullet$
E. {\it Fe production from core-collapse SN II in AGN discs.}
The observed [Fe/H] $\sim 0$ (\S\ref{sec:synopsis}-C2) corresponds to 
$Z_{\rm Fe} \sim 10^{-3}$ and a rate of Fe release into the disc
at a rate ${\dot M}_{\rm Fe} \sim Z_{\rm Fe} {\dot M}_{\rm d} \sim
10^{-3} m_8 f_\bullet M_\odot \ {\rm yr}^{-1}$ (Eqs. \ref{eq:mdotsmbh} \&
\ref{eq:mdotdisc}). The Fe is mostly released from SN II with a fractional
addition from the Fe production through the $\alpha$-chain 
reaction during the PostMS evolution.  The sub-solar
[Fe/$\alpha$] (\S\ref{sec:synopsis}-C1) is a signature of SN II ejecta. 
Scaling with an yield of $\Delta M_{\rm Fe} \sim 0.1
M_\odot$ \citep{sukhbold2016, rodriguez2021}, we infer a typical 
rate of SNII to be ${\dot N}_{\rm SNII} \sim 10^{-2} m_8 f_\bullet \ 
{\rm yr}^{-1}$ which is comparable to the rate $N_\star/\tau_\star$ 
of massive star formation inferred based on the SEPAD scenario.

\noindent
$\bullet$
F. {\it Birth of residual black holes rBHs in AGN discs.}
The collapse of stellar cores at the end of the 
PostMS evolution, leads to the formation of seed rBHs 
with a few $M_\odot$.  Their consumption 
of some residual He, $\alpha$, and Fe 
in the collapsing
cores can be used to account for the discrepancy between 
the expected conversion fraction in $Y (\simeq {\dot M}_{\rm He}/
{\dot M}_{\rm d} \sim 0.15)$ produced by the auxiliary 
stellar power sources and that observed from 
the sum of $\Delta Y \sim 0.08$ and $\Delta Z_\alpha 
\sim 0.035$ (\S\ref{sec:synopsis}-B1, Eq. \ref{eq:deltaacount}).
Since self-regulated star formation and evolution leads to
a stellar population with a top-heavy equilibrium mass function, 
the production rate of seed rBH is $\sim {\dot N}_{\rm SNII}$.  
Moreover, low-mass black holes accumulate and diffuse  
rather than migrate or flow with gas in the disc.
These seed rBHs subsequently 
gain several-fold in mass through accretion and/or 
coalescence.  These processes do not lead to further
chemical evolution due to heavy-element production, 
albeit energy dissipation during
accretion onto rBHs contributes to the auxiliary 
power.  SEPAD scenario assumes that the rBHs become the 
seeds of merging black holes, which contribute to the 
sources for intense excitation of gravitational waves.

\noindent
$\bullet$
G. {\it Multiple generations of embedded stars.}
Multiple generations of PostMS stars are needed to produce
the observed high [$\alpha$/H] (\S\ref{sec:synopsis}-A1).  Moreover, 
the synthesis of $\alpha$ elements through the triple-$\alpha$
and $\alpha$-chain reaction during the PostMS stage produces low
[N/(C+O)] ratio (\S\ref{sec:synopsis}-A2). Its redistribution 
to the observed values of [N/(C+O)] requires elemental redistribution
within the CNO cycle on the MS track of next-generation stars.   
This inference has implications on the AGN persistent time scale 
($ \gtrsim \tau_\star$ a few Myr, \S\ref{sec:sepadimp}-K) and 
the number of rBHs born in AGN discs ($\sim N_{\rm rBH} \gtrsim 
{\dot N}_{\rm SNII} \tau_\star \sim 10^4 m_8 f_\bullet$,
\S\ref{sec:sepadimp}-E \& F).

\noindent
$\bullet$
H. {\it In situ pollution in AGN discs.}
The SEPAD scenario implies the metallicity in the disc is higher than that being supplied to
the disc. Under the assumption that the NLR is 
located either beyond the outskirt of the AGN disc or in the host galaxies, the 
observed higher values of BLR's $Y_{\rm d}$ and $[\alpha$/H] than those
of the NLR (\S\ref{sec:synopsis}-B2 and A4) provide a circumstantial 
evidence for SEPAD regardless of whether gas 
in the NRL is the diluted outflow from or fresh supply to that in the BRL.
The absence of $z_\gamma$-dependence in $[\alpha$/H] and [Fe/H] 
(\S\ref{sec:synopsis}-A1 and C4) and super-solar $[\alpha$/H] with 
sub-solar [Fe/$\alpha$] (\S\ref{sec:synopsis}-C3) in AGNs' BLRs 
(in contrast to the observed $[\alpha$/H]-$z_\gamma$ and [Fe/$\alpha$]-$[\alpha$/H]
correlations for the galactic 
chemical evolution \S\ref{sec:galacticha}) also implies that they are not significantly 
affected by the enrichment or star formation activities (\S\ref{sec:synopsis}-A4) 
of their host galaxies' ISM. 

\noindent
$\bullet$
I. {\it Clearing of pollutants by accretion onto SMBHs.}
The observed $z_\gamma$-independence of $[\alpha$/H] and [Fe/H] 
(\S\ref{sec:synopsis}-A1 and C4) is in strong contrast
to galactic chemical evolution \S\ref{sec:galacticha}. 
But it is consistent
with SEPAD's inference that the pollutants released by the stars are
mixed with the disc gas, advected inward, and continuously cleared
into the central SMBHs.

\noindent
$\bullet$ 
J. {\it Self-regulated star formation rate.}
The lack of (or weak if any) $M_\bullet$ or $\lambda_\bullet$-dependence
in the observed $[\alpha$/H] (\S\ref{sec:synopsis}-A3) 
is in line with SEPAD's inference
based on the assumption of self regulate star formation rate ($\sim N_\star 
/ \tau_\star$) and capture rate \citep{artymowicz1993, goodman2003, thompson2005} in AGN discs.  
At any given radius, both the critical gas $\Sigma$ and stellar surface 
density $s_\star$ required auxiliary power $Q^+ _\star$ needed to maintain 
$Q=1$ for marginal stability increases with $M_\bullet$. In high-$\Sigma$ 
discs (with relatively high ${\dot M}_{\rm d}$, more stars form, so the 
heavy element contamination is diluted.  Thus, the $[\alpha$/H] and [Fe/H] 
dependence on $M_\bullet$ and ${\dot M}_{\rm d}$ essentially cancel with 
each other. Moreover, their dependence on the Eddington factor 
$\lambda_\bullet (\propto {\dot M}_{\rm d} /M_\bullet)$ is correspondingly weak.  

\noindent
$\bullet$ 
K. {\it AGN persistent time scale.}
The inference of multiple generations of embedded stars 
(\S\ref{sec:sepadimp}-G) imply 
that the AGN disc must have persisted for at least $\tau_\star$ (a 
few Myr). The observed sub-solar [Fe/$\alpha$] also indicates that 
SN 1a is not the predominant channel for Fe enrichment despite the 
super solar [$\alpha$/H] (\S\ref{sec:sepadimp}-E).  This 
inference implies a limit on the embedded white dwarf 
population. Since their MS progenitors have $M_\star \lesssim 
8 M_\odot$ with a lifespan $\gtrsim 5 \times 10^7$ yr and additional
time is needed for the white dwarf remnants to acquire sufficient
(Chandrasekhar) mass to undergo collapse along with SN Ia, the
sub-solar [Fe/Mg] can be used to infer an upper limit on the 
typical AGN duration ($\lesssim$ a few $10^8$ yr) and their discs
do not extend significantly beyond a few pc (i.e., the typical distance
of the ``dusty torus'') from the SMBHs.


\subsection{Some outstanding issues.}

There are several uncertainties associated with the simplest version
of the SEPAD model we have adopted in this paper.

A. {\it Stellar mass spectrum and population.} In the construction of the simplest SEPAD 
models, we assume the dominant power sources to be those MS stars with an equilibrium $m_\star$
and [He/H] of the disc gas.  We need to take into account modification due to
the $L_\star-m_\star$ for the MS stars with an evolving internal [He/H] and co-existing
PostMS stars. Supernovae and accretion onto a growing population of rBHs also contribute to
$Q^+ _\star$ and therefore the star formation rate, although they do not directly affect 
the chemical evolution of the disc.  

B. {\it Main sequence turn off.}  It is not clear whether an accretion-wind equilibrium can be stably maintained with a high retention and recycling efficiency.  The possibility of non-spherical flow and interchange 
between inflow and outflow need further theoretical analysis.

C. {\it Release of He, $\alpha$, and Fe byproducts to the disc during PostMS evolution.} Wind prescriptions 
affect PostMS-evolution models for embedded stars' interior, He-$\alpha$ 
conversion rate, $\alpha$ and Fe yield. They also lead to different values for
pre-collapse core mass, neutron stars versus rBH remnants.  The efficiency of 
the PostMS mass loss needs further studies. 

D. {\it Stellar mergers.} The coexistence of a large population (with $N_\star \gtrsim 10^3)$ of
embedded stars raise 
the possibility of their collisions and coalescence.  While the merger events would be followed
by rapid downsizing and the return to the equilibrium mass, we need to check whether
PostMS-star mergers may enhance the production and release rates of $\alpha$ and Fe 
byproducts.  

E. {\it Migration and relaxation of embedded stars.}  Although isolated embedded stars 
may migrate extensively over AGN discs, tidal interference by nearby cohorts may 
suppress this effect.  Nevertheless, we need to examine whether modifications in 
the $Q^+ _\star$ distribution may be offset by the local $Q^+ _\star \simeq Q^-$ 
thermal equilibrium in self-regulating the star formation rate.

F. {\it Radial extent of AGN discs.} The disc's mid-plane density, temperature, 
and embedded stars' accretion rate decreases with their radius.  Beyond a few pc, the
equilibrium stellar mass also decreases below those of neutron-star or white-dwarf 
progenitors, although $N_\star$ also increases with R.  Some constraints, based 
on the observed [Fe/$\alpha$] ratio, may be inferred for 
the physical extent of the star-forming region in AGN discs. 

Despite these uncertainties, 
this paper highlights the relevance and importance of AGNs abundance inference from 
their spectroscopic signatures in deciphering structure and evolution of their discs.
The SEPAD scenario also predicts the occurrence of SN II and production of rBH in 
AGN disc.  Their potential observational signatures will be explored and discussed elsewhere.

\label{lastpage}

\section*{Acknowledgements}
We thank Gary Ferland, Andy Fabian, Stan Woosley, Yixian Chen, Mohamad Ali-Deb, Andrew Cumming, 
Jamie Law-Smith, Matteo Cantiello and an anonymous referee for useful communications.

Funding for the Sloan Digital Sky Survey V has been provided by the Alfred P. Sloan Foundation, the Heising-Simons Foundation, the National Science Foundation, and the Participating Institutions. SDSS acknowledges support and resources from the Center for High-Performance Computing at the University of Utah. The SDSS web site is \url{www.sdss.org}.

SDSS is managed by the Astrophysical Research Consortium for the Participating Institutions of the SDSS Collaboration, including the Carnegie Institution for Science, Chilean National Time Allocation Committee (CNTAC) ratified researchers, the Gotham Participation Group, Harvard University, Heidelberg University, The Johns Hopkins University, L’Ecole polytechnique fédérale de Lausanne (EPFL), Leibniz-Institut für Astrophysik Potsdam (AIP), Max-Planck-Institut für Astronomie (MPIA Heidelberg), Max-Planck-Institut für Extraterrestrische Physik (MPE), Nanjing University, National Astronomical Observatories of China (NAOC), New Mexico State University, The Ohio State University, Pennsylvania State University, Smithsonian Astrophysical Observatory, Space Telescope Science Institute (STScI), the Stellar Astrophysics Participation Group, Universidad Nacional Autónoma de México, University of Arizona, University of Colorado Boulder, University of Illinois at Urbana-Champaign, University of Toronto, University of Utah, University of Virginia, and Yale University.

\section*{Data Availability}

The data analysed in this paper can be found on the Sloan Digital
Sky Survey (\url{www.sdss.org}).

\bibliography{agn_metal} 
\bibliographystyle{mn2e}
\appendix

\section {Line intensity}
\label{sec:Appendix A}

The above discussion has made use of the \nagao\ emission-line ratio as an abundance indicator.  In order to assess the robustness of this indicator, we have examined the photoionization physics responsible for the behavior of these lines.  

\cite{Nagao2006b} popularized this indicator using photoionization model predictions based on 
the “LOC model” for the BLR (see above).  However, for an examination of the physics behind the 
behavior of \civ\ and other lines, it is more practical to study the ionization structure of 
individual clouds.  Therefore, we investigated a set of Cloudy 17.01 models for single clouds 
with a slab geometry and the “Strong Bump” ionizing continuum employed by \cite{Nagao2006a}.  Our 
reference model had a hydrogen density $n_{\rm H} = 10^{10}$, an incident ionizing 
photon flux $\phi$ = $10^{19}$, a stopping column density of $10^{23}$,  and solar abundances.  The ionization parameter of this model is  $U = 10^{-0.98}$.

In the models presented here, we assume that the ionizing UV
photons are emitted from the innermost regions of the disk near the
SMBHs. Since these regions are stable against gravitational instability, the energy sources of the ionizing photon are presumably associated with accretion onto the SMBHs or the disk corona. The SEPAD
model suggests star formation occurs in gravitationally-unstable outer regions of the disk 
\citep{paczynski1978, goodman2003} beyond a few light days ($ > r_{\rm Q} \sim 0.06\ 
\alpha_\nu^{2/9} m_8 ^{7/9}$ pc for typical AGNs)
and the embedded stars sediment near the midplane \citep{artymowicz1993} where the gas density ($\rho_{\rm c} \simeq 8 \times 10^{-16} m_8 r_{\rm pc}^{-3} {\rm ~g \ cm^{-3}}$
where $r_{\rm pc} = R/$pc) is much larger than that of the interstellar medium including giant molecular clouds
and the midplane temperature $T_{\rm c} \sim 800 {m_8^{5/12} \alpha_\nu ^{-1/6} r_{\rm pc}^{-3/4}} {\rm K}$ (see \S\ref{sec:sepad0} for concept and derivation will be presented elsewhere). With nearly 
Eddington-limited luminosity, individual massive stars produce H II regions whose Strömgren radii ($R_{\rm s} \simeq 2 \times 10^{15} {r_{\rm pc}^{29/16} m_8^{-9/16}} 
{\rm cm}$)
are less than a few percent of their Roche radii $(M_\star/3 M_\bullet)^{1/3} R$ and 
the disk thickness $H=hR$ where $h\sim 0.02 r_{\rm pc}^{1/2} \alpha_\nu ^{-1/3} m_8 ^{-1/6}$). Although they provide auxiliary power to maintain hydrostatic equilibrium in the direction perpendicular to the disk plane,
their UV photons are absorbed well below the BLR, presumed to be at 
or just above the disk surface. The reprocessed radiation diffusing through the optically thick disk mainly serves to enhance the composite spectral energy distribution (SED) in the 
visual and IR continuum without significant modifications in the flux
of ionizing photons. The X-ray and UV to optical and IR luminosity
ratio is a function of the SMBH’s mass, the disks’ inner and outer radii,
the Eddington ratio and it can be, in principle, fitted to match the observed SED \citep{sirko2003}. This dependence introduces
uncertainties in deriving the metallicity and Helium abundance
directly from the equivalent width of the emission lines. However, abundance determination based on line ratios is less sensitive to the slope of the SED than the lines’ equivalent width. In order to
facilitate comparison with previous studies, we adopt \cite{Nagao2006a}’s
approximate SED in the line-ratio analysis presented here (\S\ref{sec:observations}).
 
We ran a series of models with the above parameters and a metallicity of (0.5, 1.0, 2.0, 4.0, 
and 8.0) $Z_{\odot}$.  All heavy elements were varied in proportion (including nitrogen), 
with helium held constant at ten percent of hydrogen by number of atoms.  The resulting line 
intensities are given in Table \ref{tab:Nagao_ratio}.  As C/H increases a factor of 16, the \civ\ emission line, 
relative to H$\beta$, monotonically decreases by a factor 2 (see Figure 15 of \cite{ferland1996}).
In contrast, \oiv\ remains roughly constant, and \siiv\ increases.  As a result, 
\nagao\ increases a factor of $\sim 3$.  This resembles the trend shown in 
Figure 29 of \cite{Nagao2006b}.  Between 0.5 and 8 $Z_{\odot}$, our results give a 
dependence \nagao\ $\propto Z_{\alpha}^{0.46}$, whereas Table~10 in \cite{Nagao2006a} gives 
$Z_{\alpha}^{0.44}$.  Also, at solar abundance, the ratio is 0.149 in \cite{Nagao2006a}’s model, 
smaller than our result by 0.03~dex.  This close agreement may be fortuitous, but it 
suggests that this indicator is not extremely sensitive to the BLR geometry, at least 
in the general range of traditional BLR parameters.

\setcounter{table}{0}
\renewcommand{\thetable}{A\arabic{table}}

\begin{table}
\begin{center}
\caption{Cloudy 17 results for \nagao.}
\label{tab:Nagao_ratio}
\begin{tabular}{l c c c c}
\hline
$Z_{\alpha}/Z_{\odot}$ & \civ/H$\beta$
& \oiv/H$\beta$ & \siiv/H$\beta$ & ratio
\\
\hline \hline
0.5 & 14.39 & 0.698 & 0.91 & 0.112
\\
1 & 12.72 & 0.599 & 1.35 & 0.153
\\ 
2 & 10.05 & 0.535 & 1.77 & 0.229
\\
4 & 7.86 & 0.548 & 2.02 & 0.327
\\
8 & 6.73 & 0.625 & 2.11 & 0.406
\\
\hline
\end{tabular}
\end{center}
\end{table}

The behavior of \nagao\ can be understood in a straightforward manner.  The 
intensity of an emission line is governed by the electron temperature, the abundance of 
the element , and the fractional abundance of the ion in question \citep{Osterbrock2006}.  
(However, see \cite{Temple2021} for a discussion of line saturation at high density 
and optical depth.) The collisional excitation rate is sensitive to electron temperature; 
but for the relative intensity of lines with similar ionization potentials, as here, the 
impact is modest.  The key factor is the fractional abundance of the ion, suitably averaged 
over the emitting volume.  For the same set of five Cloudy models, the fractional 
abundance of $\rm \cppp$ is  $\log X(\rm \cppp)$ = (-0.51, -0.72, -1.00, -1.28, -1.51).  (These are 
the volume-averaged values weighted by electron density.) As metallicity increases, the 
fraction of $\rm \cppp$ in the cloud decreases almost in inverse proportion to the heavy element 
abundance.  This reflects the fact that the zone in which $\rm \cppp$ predominates gets smaller 
as the carbon abundance increases.   

\setcounter{figure}{0}                     
\renewcommand\thefigure{A\arabic{figure}}

\begin{figure}
    \centering
    \includegraphics[width=0.47\textwidth]{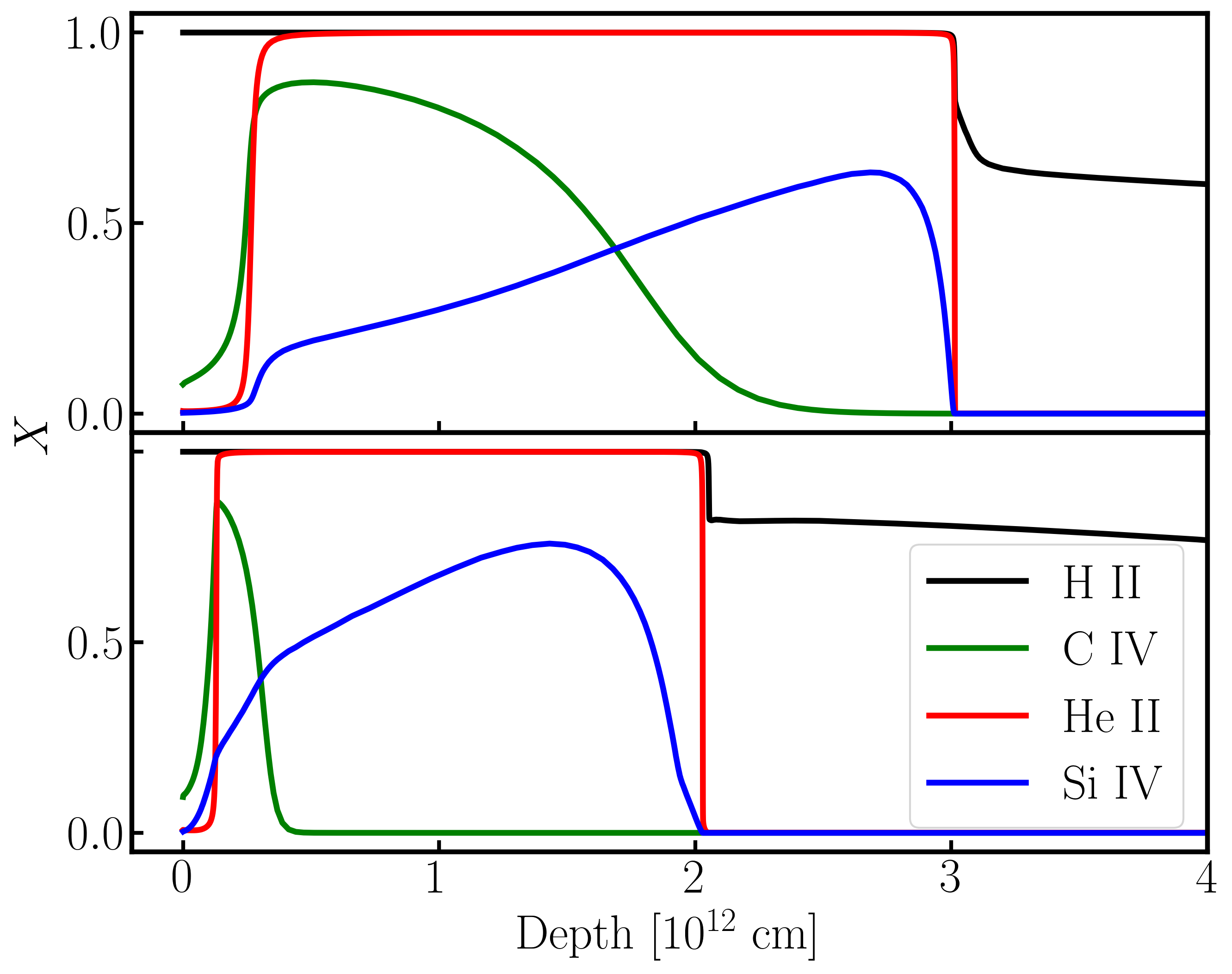}
    \caption{Cloudy results of ionization fraction, $X$, for $\rm \hp$ (black), $\rm \hep$ (red), $\rm \cppp$ (green), and $\rm \sippp$ (blue), as a function of depth. We use $\log n_{\rm H}=10$, $\log \phi=19.5$, stopping column density $\log N_{\rm H}=23~\rm cm^{-2}$, and $v_{\rm turb}=0$. For the top panel, $Z_{\alpha}=Z_{\odot}$, for the bottom panel, $Z_{\alpha}=8Z_{\odot}$. As metallicity increases, the volume of the $\rm C^{+3}$ zone decreases.}
    \label{fig:appendix_Z1}
\end{figure}

\begin{figure}
    \centering
    \includegraphics[width=0.47\textwidth]{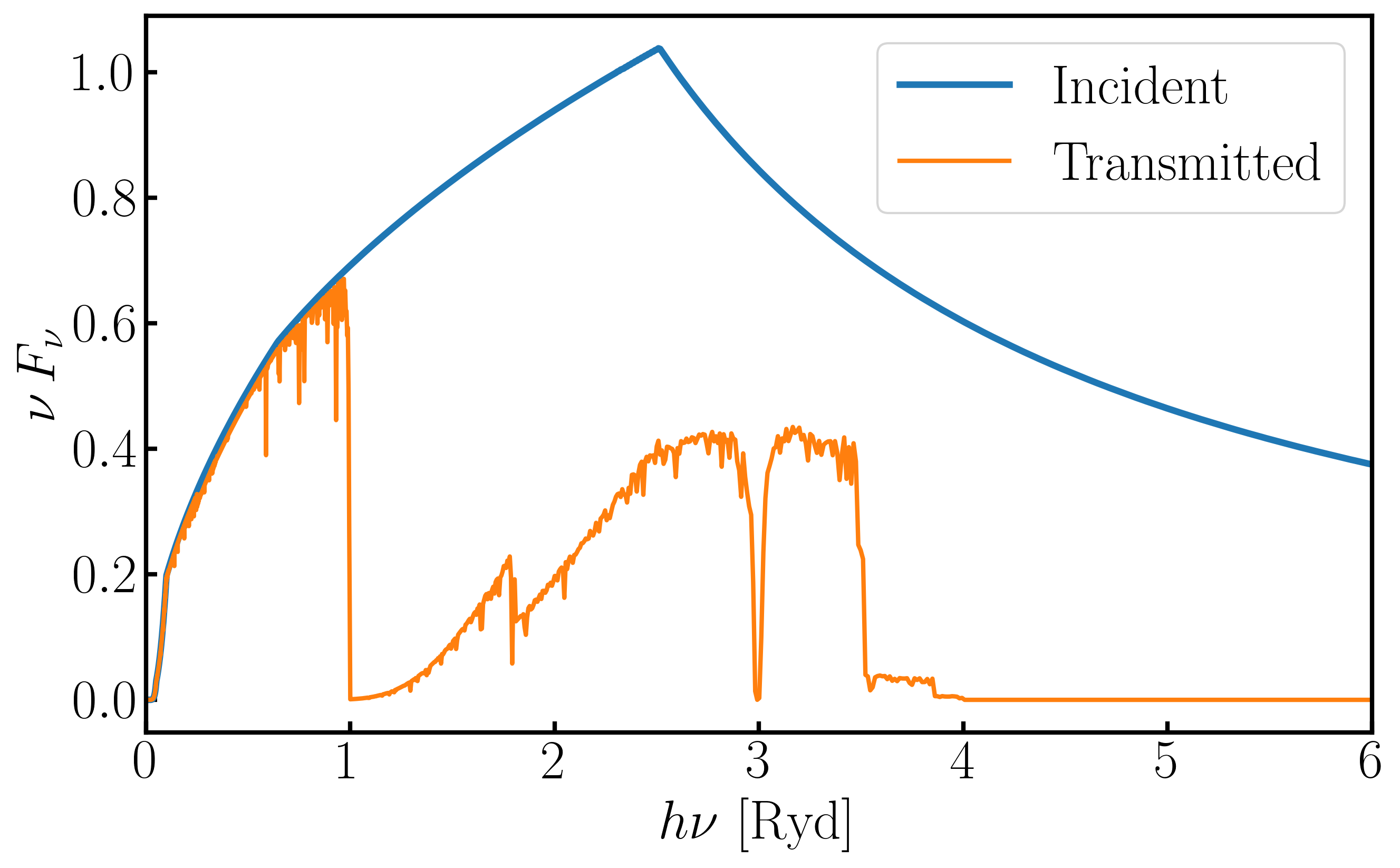}
    \caption{Incident and transmitted continuum with the same input parameter as Figure \ref{fig:appendix_Z1}, with $Z_{\alpha}=Z_{\odot}$. The transmitted continuum is computed using the slab with a cut-off at $\log d=12.26 \,\rm cm$.}
    \label{fig:appendix_continuum}
\end{figure}

Figure \ref{fig:appendix_Z1} shows the fractional abundance of several ions as a function of depth in the slab 
in the model with solar abundances.  The black curve is the $\rm \hp$ ion, red is $\rm \hep$, and green 
is $\rm \cppp$.  For the top panel, the $\rm \hp$ Strömgren sphere occurs at a depth of $3 \times 10^{12}$ cm, where the helium also goes 
neutral.  The $\rm \hepp$ Strömgren length occurs at $2 \times 10^{11}$cm, where the $\rm \hep$ abundance rises 
abruptly, as does $\rm \cppp$.  However, the  $\rm \cppp$ ion gives way to  $\rm \cpp$ at a depth of $1.5 \times 10^{12}$ cm, 
only half the depth of the hydrogen Strömgren sphere.  This results from exhaustion of 
the available photons with energy above 3.52 Rydbergs that are capable of ionizing $\rm \cpp$.  The bottom panel shows the 
ionization fractions versus depth for a metal abundance 8 times solar (other parameters 
unchanged).  Now the $\rm \cppp$ zone is a tiny fraction of the volume of the $\rm \hep$ zone, having 
an effective width of only about $2 \times 10^{11}$ cm.
Figure \ref{fig:appendix_continuum} shows the ionizing continuum $\nu F_{\nu}$ as a function of frequency in Rydbergs.  
The black curve is the incident continuum, and the red curve is the transmitted continuum 
at the outer edge of the $\rm \cppp$ zone (a depth of $1.8\times 10^{12} \rm \, cm$).  Here the continuum is completely 
extinguished above 4.0 Ryd, and nearly extinguished between 3.5 and 4 Ryd.  This is the 
reason that the $\rm \cppp$ abundance drops as this depth is approached.   

Essentially, the $\rm \cppp$ ion forms its own Strömgren sphere well inside the outer boundary 
of the $\rm \hep$ zone. The volume of the $\rm \cppp$ zone is, to a first approximation, determined 
by the familiar ``photon counting argument’’ \citep{Osterbrock2006}.  As the metal 
abundance goes up, the recombination rate per unit volume for $\rm \cppp$ goes up proportionally; 
and every recombination is balanced by a photoionization that consumes a photon between 
3.52 and 4.0 Ryd. Thus, the number of $\rm \cppp$ ions in the slab remains roughy constant.  
Given that the electron temperature decreases with increasing metallicity, the intensity 
of the \civ\ emission line actually decreases as the metal abundance increases.  In 
contrast, the ionization fraction of $\rm Si^{+3}$ remains nearly constant at $\log X = -0.75$
as metallicity increases.  The ionization potential of $\rm Si^{+2}$ is 2.46 Ryd, so the 
relatively unabsorbed continuum between 2.5 and 3.5 Ryd is available to maintain the 
abundance of $\rm Si^{+3}$.  Therefore, the increasing elemental abundance of silicon gives an 
increasing intensity of the \siiv\ emission line. 

The straightforward physics behind \nagao\ gives confidence in this indicator of 
metallicity, at least in a qualitative sense.  But what factors may affect the 
calibration of the line ratio in terms of absolute abundances?  One issue is the 
``turbulence'' within the BLR.  We have mentioned above the importance of this 
parameter for the \feii\ intensity and other predictions of Cloudy models 
\citep{sarkar2021, Temple2021}.  The turbulence parameter $v_{\rm turb}$ has some effect 
on the \nagao\ ratio.  For our $\log n_{\rm H} = 10$, log $\phi$ = 19.5 model 
with solar abundances, this ratio drops from 0.15 for zero turbulence to 0.11 
for  $ v_{\rm turb} = 10^7 $ cm s$^{-1}$, then rises slightly to $0.12$ for $5 \times 10^7$ cm s$^{-1}$.  The effect of 
this is to increase the metallicity required to reproduce a particular value 
of \nagao\ when $v_{\rm turb} \approx 10^7$cm s$^{-1}$, compared to $v_{\rm turb} = 0$.

The nitrogen abundance has some impact on \nagao. The ionization potential of $\rm \npp$
is 3.49 Ryd, nearly the same as for $\rm \npp$.  Therefore, photoionization of $\rm \npp$ competes 
for the same scarce photons between 3.5 and 4.0 Ryd that ionize $\rm \cpp$.  A high abundance 
of nitrogen will therefore reduce the volume in which $\rm \cppp$ is abundant.  To explore 
this effect, we reran the above model with $Z = 4 Z_\odot$ but with N/O = 4(N/O)$_\odot$ (i.e., 
``secondary’’ nitrogen). The increase in nitrogen reduces the intensity of \civ\ 
from 7.9 to 5.5, reflecting a decrease of 0.22 dex in the fractional abundance of 
$\rm \cppp$. The \nagao\ ratio increases from 0.33 to 0.39. This mimics the effect of 
raising the metallicity from $Z_{\alpha} = 4Z_\odot$ and 8 $Z_\odot$ for solar N/O. 
Therefore, a simultaneous determination of the nitrogen abundance will benefit 
the accuracy of the metallicity derived from the \nagao\ ratio.
The effect of helium abundance on \nagao\ is less significant. In the preceding
model, with N/O = 4(N/O)$_\odot$ but with He/H increased to 0.175 rather than 0.10, the \nagao\ ratio drops only from 0.39 to 0.36.

In this paper, we have focused on the \nagao\ indicator for the abundance of the alpha-elements.  Other indicators can give substantially different results for the BLR metallicity. For example, comparison of Figures 22 and 29 of \cite{Nagao2006b},
for the more luminous quasars, gives roughly solar metallicity from  \ciiiciv\ but nearly 
$10Z_{\odot}$ from Al~{\sc iii}/\civ\ and even higher for the \nv\ indicators.  However, as noted by \cite{Lai2022}, 
Al~{\sc iii} and \nv\ are more difficult to measure, and 
\ciii\ is sensitive to the ionization parameter and to 
collisional de-excitation. Moreover, many of the line ratios discussed by \cite{Nagao2006a} and others 
use \civ\ as the denominator. We have argued here that the ionization structure of \civ\ largely 
drives the metallicity dependence of these ratios, and therefore they cannot be considered independent.

\section{Galactic chemical evolution versus SEPAD} 
\label{sec:galacticha}

For the purpose of constraining the SEPAD scenario (\S\ref{sec:sepad0}, \S\ref{sec:sepadimp})
with the abundance properties inferred from observations (\S\ref{sec:synopsis}), 
we highlight how different physical processes lead to diverse 
paths of chemical evolution in the AGN environment
versus that in the Galaxy. 

{\it A. Galactic stars'} (including the Sun) elemental distribution is the byproduct of their
chemical evolution with a conventional initial mass function (IMF) and their 
mass-dependent evolutionary tracks in the Milkyway disc,
halo\citep{venn2004}, and satellite dwarf spheroidal galaxies 
\citep{shetrone1998, hill2019} in the Galatcic halo.  

\noindent
$\bullet$ A1. {\it Enrichment of He, $\alpha$, and Fe abundances.}
Hydrogen (H) is converted into helium (He) through the pp chain or
the CNO cycle on the Main Sequence (MS) while $\alpha$ (including C, O, Si, Mg) 
and some Fe elements are produced through the triple-$\alpha$ and 
$\alpha$-chain process during the post Main Sequence (PostMS) evolution. 
Nitrogen (N) is a secondary element enhanced through the CNO cycle
on the MS without changes in the total C+N+O abundance.
High and intermediate-mass stars ($\gtrsim 8 M_\odot$), 
detached from their natal molecular clouds, rapidly evolve off (in 
$\lesssim 10^8$ yr) their MS track, release He and CNO-enriched winds 
during the Wolf-Rayet and PostMS red giant phases, respectively.

\noindent
$\bullet$ A2. {\it The {\rm [$\alpha$/Fe]} ratio.}
Although Mg is produced in these stars' deep interior during their hydrostatic phase
\citep{weaver1978, nomoto1988}, its ejection is delayed\citep{choi2016} until the 
advanced PostMS or type II supernova (SN II) phase, accompanied by comparable amounts 
of Fe and Si and larger masses of C and O \citep{thielemann1996}.  
Since both $Z_\alpha$ and $Z_{\rm Fe}$ of the  ejecta are several times larger than those inferred for 
zero-age-main-sequence (ZAM) Sun,  this feedback process
increases these abundances towards the solar values in the initially pristine ISM.  Nevertheless, 
the abundance ratio of $Z_{\rm Fe}/Z_\alpha$ is below its solar value due to the 
slightly less efficient production of Fe than $\alpha$ elements in massive stars
\citep{woosley2002}.  Less massive ($ \leq 8 M_\odot$) stars take longer ($\gtrsim 10^8$ yr) 
time to evolve through the MS phase, becoming red giants with N-enriched winds and ultimately 
white dwarfs. Those compact remnants located 
in suitable binary systems eventually undergo type Ia supernova (SN Ia) explosions, with the ejection 
of comparable masses of CNO ($\sim 0.2 M_\odot$) and Si but larger amounts of Fe ($\sim 0.6 
M_\odot$) \citep{nomoto1984, iwamoto1999}.

These diverse yields from SN Ia and SN II 
are reflected in the differences in [$\alpha$/Fe] for stars formed in different epochs.
Among the Galactic halo population II and disc stars as well as 
stars in the Local Group dwarf spheroidal galaxies, the observed values of [$\alpha$/Fe], 
including [Mg/Fe], are elevated by $\sim 0.3 - 0.5$~dex (within $\sim 2 \sigma$ standard 
deviation), while both [$\alpha$/H] and [Fe/H] are down by 2 to 3 dex\citep{tolstoy2009} over an 
age span of several Gyr. 
This pattern implies that the $\alpha$ 
elements contained in these stars, in particular, O, are primarily produced by 
an early generation of massive stars, whereas the Fe-peak elements in later-generation
Pop I stars are mostly generated by SN Ia
from modest-mass ($\lesssim 8 M_\odot$) stars that 
take longer to evolve \citep{matteucci2003}. Nevertheless, both the observed Pop II abundances 
and the theoretical yields imply that a value of [Fe/Mg]$_{\rm II}$ 
(=[Fe/$\alpha$]$_{\rm II} \sim -0.3$), 
not vastly lower than the solar value can 
be generated by SN II alone \citep{heger2002}. In the ramp-up to the 
solar value of [Fe/H] and [Fe/Mg] (=[Fe/$\alpha]$) over several Gyr (at cosmic 
red-shift $z_\gamma \simeq 2-3$), SN Ia contribute 
more Fe than do type II supernovae for a population of stars. 

\noindent
$\bullet$ A3. {\it He enrichment.}
The MS stars convert H into He and some of which is returned 
to the ISM through WR and PostMS winds as well as SNs, along with the $\alpha$ and Fe byproducts.
Galactic chemical evolution models, studies of H II regions in the Milky Way, 
and measurements for individual stars in the solar neighborhood \citep{carigi2008, peimbert2016, valerdi2021}
indicate changes in the 
Galactic He mass-fraction abundance from its values $Y_0 (\sim 0.245)$
to its present-day value ($Y_\odot \simeq 0.28$)
\begin{equation}
\Delta Y_{\rm G} \simeq 3.3 \Delta Z_{\rm O, G} \simeq 1.8 \Delta Z_{\rm \alpha, G}
\label{eq:deltaydeltaz}
\end{equation}
where $\Delta Z_{\rm O, G}$ and $\Delta Z_{\rm \alpha, G}$ 
are the average changes in the 
mass fraction of oxygen and $\alpha$ elements in the Galaxy ($\simeq Z_{\rm O, G}$ and
$\simeq Z_{\rm \alpha, G}$ their present-day values).

\noindent
$\bullet$ A4. {\it Secondary production of N.}
Primary-generation stars with $Z_{\rm \alpha, G} \lesssim Z_{\odot} /5$ in the Galactic 
halo and nearby dwarf galaxies have nearly constant $Z_{\rm N, p}/Z_{\rm O, p} (\sim 10^{-1.6})$ 
whereas this ratio steeply increases with $Z_{\rm \alpha, G}$ among the more metal-rich 
younger stars\citep{edmunds1978, 
henry2000}.  This difference has been attributed to a transition between the primary/secondary N production 
in which 
\begin{equation}
[{\rm N / H}]_{\rm p} \simeq [{\rm \alpha/H}]_{\rm p}, \ \ \ 
[{\rm N / H}]_{\rm s} \simeq 2 [{\rm \alpha/H}]_{\rm s}, \ \ \ 
    \label{eq:nabundance}
\end{equation}
for $[{\rm \alpha/H}]_{\rm G} \lessgtr -0.7$ respectively during the 
earliest/later phases of Galactic chemical evolution \citep{
arnett1996, meynet2002}.  
 
{\it B. SEPAD's stellar populations.} They form in or captured by marginally stable AGN discs(\S\ref{sec:sepad0}). They continue to grow until they attain 
an accretion-wind equilibrium with a radial-dependent, skewed top-heavy IMF.

\noindent
$\bullet$ B1. {\it Hypothetical long-lasting massive MS stars} 
accrete disc gas, convert H into He through CNO cycle,
release He-enriched byproducts, and substantially increases $Y_{\rm d}$ 
and N/(C+O) without changing $Z_\alpha$ and $Z_{\rm Fe}$ \citep{cantiello2021}. 

\noindent 
$\bullet$ B2. {\it Transition to PostMS evolution} require high-degree of 
He-byproduct recycling (Ali-Deb et al, submitted).  He retention also
increases $\mu_\star$ and reduces equilibrium mass by releasing 
He-byproducts to the disc.

\noindent 
$\bullet$ B3. {\it On the PostMS track,}
He is converted into $\alpha$ elements through 
triple-$\alpha$ and $\alpha$-chain reactions and release intense
wind with $\alpha$-enriched byproducts. 

\noindent 
$\bullet$ B4. {\it Insufficient PostMS, pre-collapse 
mass loss} would lead to the direct collapse and the formation 
of relatively massive ($\sim 10^{2} M_\odot$) residual black 
holes without SN II. The $\alpha$ and Fe injection into the 
disc would be largely suppressed.  But, if their pre-collapse 
mass is able to reduce to $\sim 8-15 M_\odot$ by strong PostMS winds, 
these stars would undergo 
SN II and release $\alpha$ and Fe-peak ejecta with a sub-solar
[Fe/$\alpha$], and leave behind remnant black holes 
(with a few $M_\odot$) and neutron stars.

\noindent 
$\bullet$ B5. {\it Embedded residual black holes} continue to gain mass
without heavy-element pollution through gas accretion and merger 
events which also lead to LIGO/Virgo gravitational wave events.  

\noindent 
$\bullet$ B6. {\it Neutron stars and white dwarfs} may formed in the outer region
of persistent ($\gtrsim 10^8$ yr) low-luminosity AGN discs around modest
SMBH. Merging neutron stars lead to gravitational wave events with copious 
r-processes byproducts (such as GW170817) and accretion of disc gas onto white
dwarfs lead to their collapse with SN 1a and super solar [Fe/$\alpha$] ejecta.

\noindent 
$\bullet$ B7. {\it AGN discs} are supplied at their outer radii by 
metal-deficient gas (with [$\alpha/H$] and [Fe/H] $\lesssim 0$) from their host galaxies.
The injected He, $\alpha$, and Fe byproducts (at large to small disc radii) 
are mixed with the turbulent disc gas as it flows inwards (due to viscous
stress) towards the SMBH. An asymptotic equilibrium between injection, diffusive mixing, 
and advection is established with a radial distribution of these elements
(Zhou et al, in preparation).

\end{document}